\documentclass[article,journal]{IEEEtran}

\usepackage{cite}
\usepackage{amsmath}
\usepackage{amsthm}
\usepackage{color}
\usepackage{graphicx}
\usepackage[usenames,dvipsnames]{pstricks}
\usepackage{epsfig}
\usepackage{pst-grad} % For gradients
\usepackage{pst-plot} % For axes
\usepackage{upgreek} 
\usepackage{mathrsfs}
\usepackage{amssymb}
\usepackage{psfrag}
\usepackage{bbm}
\usepackage{xspace}

\usepackage{calc}
\usepackage{graphicx}
\graphicspath{{figs/}{../figs/}}  

\theoremstyle{plain}

\newtheorem{theorem}{Theorem}
\newtheorem{lemma}{Lemma}
\newtheorem{corollary}{Corollary}

\theoremstyle{definition}
\newtheorem{remark}{Remark}
\newtheorem{example}{Example}

\input{mathlig}
\usepackage{xspace}
\usepackage{bbm}
\input{mathlig}

\newcommand{\muspace}{\mspace{1mu}}

\DeclareRobustCommand{\scond}{\mathchoice{\muspace\vert\muspace}{\vert}{\vert}{\vert}}
\mathlig{|}{\scond}

\DeclareRobustCommand{\discint}{\mathchoice{\mspace{-1.5mu}:\mspace{-1.5mu}}{\mspace{-1.5mu}:\mspace{-1.5mu}}{:}{:}}
\mathlig{::}{\discint}
\newcommand{\suchthat}{\colon}

\newcommand{\Cc}{\mathcal{C}}

\newcommand{\Ec}{\mathcal{E}}

\newcommand{\Jc}{\mathcal{J}}

\newcommand{\Sc}{\mathcal{S}}
\newcommand{\Tc}{\mathcal{T}}

\newcommand{\Xc}{\mathcal{X}}
\newcommand{\Yc}{\mathcal{Y}}

\newcommand{\Cr}{\mathscr{C}}

\newcommand{\gv}{{\bf g}}

\newcommand{\xv}{{\bf x}}
\newcommand{\yv}{{\bf y}}

\newcommand{\Wh}{{\hat{W}}}

\newcommand{\mh}{{\hat{m}}}

\newcommand{\wh}{{\hat{w}}}

\def\a{\alpha}
\def\b{\beta}
\def\g{\gamma}
\def\d{\delta}
\def\e{\epsilon}

\let\P\relax
\DeclareMathOperator\P{\textsf{P}}

\def\error{\mathrm{e}}

\newcommand{\U}{\mathrm{Unif}}

\def\textiid{i.i.d.\@\xspace}
\newcommand\iid{\ifmmode\text{ i.i.d. } \else \textiid \fi}

\newcommand{\Ff}{\mathbb{F}}

\def\clap#1{\hbox to 0pt{\hss#1\hss}}
\def\mathclap{\mathpalette\mathclapinternal}
\def\mathclapinternal#1#2{%
  \clap{$\mathsurround=0pt#1{#2}$}}

\let\oldstackrel\stackrel
\renewcommand{\stackrel}[2]{\oldstackrel{\mathclap{#1}}{#2}}

\begin{document}

\title{Quickest Sequence Phase Detection}

\author{Lele Wang~\IEEEmembership{Member,~IEEE}, Sihuang Hu~\IEEEmembership{Member,~IEEE}, and Ofer Shayevitz~\IEEEmembership{Senior Member,~IEEE}
\thanks{This work was supported by an ERC grant no. 639573, and an ISF grant no. 1367/14. The material in this paper was presented in part in the IEEE International Symposium on Information Theory 2016, Barcelona, Spain.}
\thanks{L. Wang is jointly with the Department of Electrical Engineering, Stanford University, Stanford, CA, USA and the Department of Electrical Engineering - Systems, Tel Aviv University, Tel Aviv, Israel (email: wanglele@stanford.edu).}
\thanks{S. Hu and O. Shayevitz are with the Department of Electrical Engineering - Systems, Tel Aviv University, Tel Aviv, Israel (emails: sihuanghu@post.tau.ac.il, ofersha@eng.tau.ac.il).}}

\parskip 3pt

\maketitle

\begin{abstract}
A phase detection sequence is a length-$n$ cyclic sequence, such that the location of any length-$k$ contiguous subsequence can be determined from a noisy observation of that subsequence. In this paper, we derive bounds on the minimal possible $k$ in the limit of $n\to\infty$, and describe some sequence constructions. We further consider multiple phase detection sequences, where the location of any length-$k$ contiguous subsequence of each sequence can be determined simultaneously from a noisy mixture of those subsequences. We study the optimal trade-offs between the lengths of the sequences, and describe some sequence constructions. We compare these phase detection problems to their natural channel coding counterparts, and show a strict separation between the fundamental limits in the multiple sequence case. Both adversarial and probabilistic noise models are addressed.
\end{abstract}

\thispagestyle{empty}

\section{Introduction}
\label{sec:intro}

A magician enters the room with a 32-card deck. He invites five volunteers to the stage and claims he will  read their minds. Another volunteer is asked to cut the deck a few times and pass the top five cards to the volunteers, one for each. ``Now I need you to think about your card and I will tell what it is,'' the magician says. Silence. ``Please concentrate! Think harder.'' A long pause. ``Okay, the weather is not good today. It is interfering with the brainwaves between us. I need you to work with me a bit,'' the magician begs. ``Could the people with red cards move one step closer to me?'' Another long pause. ``Hmm, you have the six of clubs. You have the five of spades...'' Sure enough, he gets them all!

This is Diaconis' mind-reading trick~\cite{Diaconis--Graham2011,goresky2012algebraic}. The magic makes use of a binary de Bruijn sequence of order 5~\cite{deBruijn1946}, which is a length-32 circulant binary sequence such that every length-5 binary string occurs as a contiguous subsequence exactly once. The magician enters the room with the 32 cards prearranged such that their color (black/red) corresponds to the de Bruijn sequence. Cutting the deck only shifts the sequence cyclically. By the property of de Bruijn sequence, knowing the colors reveals the location (or \textit{phase}) of the 5 contiguous cards inside the deck, hence uniquely determines their identities. More generally, this trick can be performed with $k$ volunteers and a deck of size $n=2^k$, by using a de Bruijn sequence of order $k$, which is a binary sequence such that every length-$k$ binary string occurs as a contiguous subsequence exactly once~\cite{deBruijn1946}.  

Suppose now that some of the volunteers are not collaborative and may lie when asked about their card color. Can the magician still guess the cards correctly? In other words, can one design a length-$n$ sequence such that the set of all length-$k$ contiguous subsequences forms a good error-correcting code? Besides its appeal as a card trick, such a sequence can also be useful e.g. for phase detection in positioning systems. Imagine 
that a satellite sends the length-$n$ sequence periodically. A user hearing a noisy chunk of the sequence would like to figure out the location of his chunk within the original sequence, so as to measure the transmission delay and compute his distance to the satellite. Fixing the sequence length $n$ (which results in a given ambiguity of the distance estimation), it is clearly desirable to minimize $k$, as this results in the fastest positioning. Clearly, $k$ cannot be smaller than $\log{n}$, and this lower bound can be achieved in case there is no noise, by using a de Bruijn sequence of order $k$. As we shall see, in the noisy case $k=O(\log n)$ is also sufficient, and we will in fact be interested in characterizing the exact constant $\frac{\log{n}}{k}$, which will be referred to as \textit{rate}. 

In reality, positioning systems typically employ multiple satellites, each transmitting its own length-$n_i$ sequence. Sequences get combined through a multiple access channel (MAC) when reaching the user. Upon hearing a length-$k$ chunk of the combined sequence, the user wishes to measure his distance to all of the satellites by locating the chunk within each one of the sequences. We note that existing techniques (such as GPS~\cite{wiki:gps}) typically employ sequences (e.g. Gold codes~\cite{Gold1967}) that possess good autocorrelation and cross-correlation properties, and use $k = N\cdot n_1 = \cdots = N\cdot n_L$, for some repetition factor  $N\geq 1$. From our perspective, these systems hence operate at zero rates. In fact, when the repetition factor $N>1$, this does not precisely fall under our setup; we further remark on this in Example~\ref{ex:GPS}. In what follows, we focus on fast positioning at non-zero rates. We are interested in characterizing the optimal trade-offs among 
$
 \bigl(\frac{\log{n_1}}{k},\cdots,\frac{\log{n_L}}{k}\bigr)
$
that ensure successful detection, as well as in constructing sequences that achieve the optimal trade-offs. 

In what follows, we refer to the first problem, which only involves a single-sequence design, as \emph{point-to-point phase detection}. We refer to the second problem as \emph{multiple access phase detection}. Different noise models are considered: the adversarial noise and the probabilistic noise. For the probabilistic noise, different error criteria are discussed: the vanishing error criterion and the zero error criterion. These models are defined formally in the sequel. We also compare the phase detection problems to their natural channel coding counterparts.

\subsection{Point-to-Point Phase Detection}
\label{sec:intro-p2p}

In Sections~\ref{sec:advs},~\ref{sec:prob},~and~\ref{sec:prob-zero}, we consider point-to-point phase detection. 

An $(n,k)$ \emph{point-to-point phase detection scheme} consists of
\begin{itemize}
 \item a sequence $x^n \triangleq (x_1, x_2,\ldots,x_{n}) \in \Xc^n$, and

 \item a detector $\mh \suchthat \mathcal{Y}^k \to [n]\cup \mathrm{e}$, where $[n]\triangleq\{1,2,\ldots,n\}$ and $\mathrm{e}$ is an error symbol. 
\end{itemize}

We assume that the detector observes a noisy version $y^k$ of the sequence $x_m^{m+k-1}$, and attempts to correctly identify the \emph{phase} $m$.   Clearly, any reliable scheme would require $k \ge \log_{|\Xc|}n$. Thus, it is natural to define the \emph{efficiency} of  a scheme as the excess multiplicative factor it uses over the minimal possible, i.e., $k/\log_{|\Xc|}{n}$. However, for comparison to channel coding, it would be more convenient to work with the inverse of this quantity and take logarithms in base 2, namely work with the \emph{rate}
\[
R \triangleq \frac{\log_2{n}}{k}.
\]
We note that any phase detection scheme induces a \textit{codebook}\footnote{The codebook is treated as a multiset, namely there might be repetitions in its elements.} $\mathcal{C} = \{x_{m}^{m+k-1}\suchthat m\in[n]\}\subseteq \Xc^k$ of rate $R$. Here and throughout indices are taken cyclically, modulo the set $[n]$. Also, we assume throughout that $k\leq n$.

We discuss three distinct models: the adversarial noise model in Section~\ref{sec:advs}, the probabilistic noise with vanishing error in Section~\ref{sec:prob}, and the probabilistic noise with zero error in Section~\ref{sec:prob-zero}. For convenience, let the function $\phi(m; x^n)$ return the length-$k$ contiguous subsequence of $x^n$ starting at phase $m$, i.e., $\phi(m; x^n) = x_{m}^{m+k-1}$. We will typically omit the dependence on the sequence $x^n$, and simple write $\phi(m; x^n) = \phi(m)$. 

For the adversarial noise model, we assume that $\Xc = \Yc = \{0,1\}$ and the observation sequence $y^k$ is obtained from $\phi(m)$ by flipping at most $pk$ bits, where $m$ is the correct phase, and $p$ is fixed and given. We define the \emph{minimum distance} of a scheme as the minimum Hamming distance of its induced codebook. A rate $R$ is said to be \emph{achievable} in this setting if, for a divergent sequence of $k$'s, there exist $(n,k)$ schemes with $\frac{\log{n}}{k} \ge R$, such that $m$ can be recovered from $y^k$ without error. Namely, we require the scheme to have a minimum distance $d > 2pk$. The \emph{capacity} of adversarial phase detection $C_\text{ad}(p)$ is defined as the supremum over all achievable rates\footnote{Here we define capacity asymptotically. Note that similarly to adversarial channel coding, it is not guaranteed short sequences with rate above the capacity do not exist.}. 

Several works have addressed this noise model in the literature. The trade-off between the rate and the minimum distance of the code was studied in~\cite{Kumar--Wei1992, Hagita--Matsumoto--Natsu--Ohtsuka2008}. Kumar and Wei provided a lower bound on $d$ in the regime of $d \le \sqrt{k}$ for \emph{$m$-sequences}, which are generated by linear feedback shift registers~\cite{Kumar--Wei1992}. Some explicit sequence constructions were also provided in~\cite{Krishnamachari--Yedavalli2007, Malzbender--Reth--Ordentlich2012, Jorissen--Maesen--Doshi--Bekaert2014, Horvath--Herout--Szentrandrasi--Zacharias2013,Berkowitz--Kopparty2015}. 
By a concatenation of an optimal binary channel code with the Reed--Solomon code, Berkowitz and Kopparty have recently constructed a phase detection scheme with nonzero rate and nonzero relative distance~\cite{Berkowitz--Kopparty2015}. For generalization to two dimensional phase detection, see~\cite{MacWilliams--Sloane1976, Etzion1988, Paterson1994, Bruckstein--Etzion--Giryes--Gordon--Holt--Shuldiner2012}.

In Section~\ref{sec:advs}, we focus on the tradeoff between the rate and the minimum distance in the asymptotic limit. We note that a codebook induced by any phase detection scheme can be used as a channel code in the standard binary adversarial channel model~\cite{Hamming1950}. The capacity of the latter setup is unknown. Clearly however, any upper bound for that capacity, such as the MRRW upper bound~\cite{McEliece--Rodemich--Rumsey--Welch1977}, also serves as an upper bound for $C_\text{ad}(p)$. The best known binary adversarial channel coding lower bound is given by Gilbert and Varshamov~\cite{Gilbert1952,Varshamov1957}. Applying the Lov\'{a}sz local lemma~\cite{Lovasz--Erdos1975}, we show 
in Section~\ref{sec:advs-limit} that this rate is also achievable for adversarial phase detection. In Section~\ref{sec:advs-code}, we characterize the family of \emph{linear} phase detection schemes and study their performance.

For the probabilistic noise model with vanishing error criterion, we assume that the phase is uniformly distributed, i.e., $M \sim \mathrm{Unif}[n]$. We further assume that the noisy observation $y^k$ is obtained from $\phi(m)$ via a discrete memoryless channel $p(y|x)$. The probability of error is defined as
\[
 P_e^{(k)} = \P\{M \neq \hat{m}(Y^k)\}.
\]
A rate $R$ is said to be \emph{achievable} if, for a divergent sequence of $k$'s, there exist $(n,k)$ schemes with $\frac{\log{n}}{k} \ge R$ and $\lim_{k\to \infty}P_e^{(k)} = 0$. The \emph{vanishing error capacity} of probabilistic phase detection $C_\text{ve}$ is defined as the supremum over all achievable rates.

As before, the codebook induced by any phase detection scheme is also a channel code. Thus, the Shannon capacity of the channel $p(y|x)$ is an upper bound for $C_\text{ve}$. In Section~\ref{sec:prob-limit}, we show that in fact $C_\text{ve}$ equals the Shannon capacity. Moreover, we present in Section~\ref{sec:prob-code} a concatenated construction with $O(k\log k)$ complexity that achieves the capacity of probabilistic phase detection. As a consequence, this construction also establishes the equivalence between channel coding and phase detection for this noise model.

For the probabilistic noise model with zero error criterion, we again assume that the noisy observation $y^k$ is obtained from $\phi(m)$ via a discrete memoryless channel $p(y|x)$. A rate $R$ is said to be \emph{achievable} if, for a divergent sequence of $k$'s, there exist $(n,k)$ schemes with $\frac{\log{n}}{k} \ge R$ such that the phase $m$ can be recovered with \emph{zero error} for any $m \in [n]$. Similar to Shannon's zero error channel coding~\cite{Shannon1956}, achievable rates can be
equivalently defined on the \emph{confusion graph} $G = (\Xc,E)$ associated with the channel $p(y|x)$. Here the vertex set is $\Xc$ and two distinct vertices are connected $(u,v) \in E$  if they may result in the same output, i.e., there exists a $y \in \Yc$ such that $p_{Y|X}(y|u) > 0$ and $p_{Y|X}(y|v)>0$. Let $G^k = (\Xc^k,E_k)$ be the $k$-fold \emph{strong product} of $G$, where two distinct vertices are connected $(u^k, v^k) \in E_k$ if for all $i \in [k]$, either $u_i = v_i$ or $(u_i, v_i)
\in E$. Then, a rate $R$ is achievable if and only if, for a divergent sequence of $k$'s, there exist $(n,k)$ schemes with $\frac{\log{n}}{k} \ge R$ such that $(\phi(m), \phi(m')) \notin E_k$ for any two distinct phases $m, m'\in [n]$, or in other words, the induced codebook forms an independent set of $G^k$. The \emph{zero error capacity} $C_\text{ze}(G)$ is defined as the supremum over all achievable rates. %{\color{red}Despite the fact that the zero error channel capacity is generally unknown, we show in Section~\ref{sec:prob-zero} that the zero error capacity for phase detection coincides with its channel coding counterpart.}

%{\color{red}
%We note the distinction between phase detection and channel coding. For the channel coding, if a rate $R$ is achievable at some length $k$, it is also achievable for all multiples of $k$ (by concatenation) and thus for a divergent sequence of $k$'s. However, this argument cannot be applied to the phase detection setting, since concatenating the codewords of two induced codebooks may not necessarily result in a new codebook that can be chained up into a single sequence.
%}
We note the distinction between phase detection and channel coding under the zero error criterion. For zero error channel coding (in contrast to vanishing error and adversarial channel coding) if a rate $R$ is achievable at some length $k$, it is also achievable for all multiples of $k$ (by concatenation) and thus for a divergent sequence of $k$'s. However, this argument cannot be applied to the phase detection setting, since concatenating the codewords of two induced codebooks may not necessarily result in a new codebook that can be chained up into a single sequence. Nevertheless, and despite the fact that the zero error channel capacity is generally unknown, we show in Section~\ref{sec:prob-zero} that the zero error capacity for phase detection coincides with its channel coding counterpart.  

\subsection{Multiple Access Phase Detection}

In Sections~\ref{sec:mac-prob}~and~\ref{sec:mac-advs}, we consider multiple access phase detection. We only discuss the two-user case for simplicity. But all the results extend to more users. 

An $(n_1,n_2,k)$ \emph{multiple access phase detection scheme} consists of
\begin{itemize}
 \item two sequences $x_1^{n_1} = (x_{11}, x_{12},\ldots,x_{1,n_1}) \in \Xc_1^n$ and $x_{2}^{n_2} = (x_{21},x_{22},\ldots,x_{2,n_2}) \in \Xc_2^n$, and
 \item a detector that declares two phase estimates $\mh_1\suchthat \Yc^k \to [n_1]\cup \{\error\}$ and $\mh_2\suchthat \Yc^k \to [n_2]\cup \{\error\}$.
%  an estimate $(\mh_1,\mh_2) \in [n_1]\times[n_2]$ or an error $\mathrm{e}$ to each observation $y^k$, where $[n]\triangleq\{1,2,\ldots,n\}$. 
\end{itemize}

We assume that the detector observes $y^k$, which is the output of a discrete memoryless multiple access channel $(\mathcal{X}_1\times \mathcal{X}_2, p(y|x_1,x_2), \mathcal{Y})$ with the two inputs $\phi_1(m_1) = \phi_1(m_1; x_1^{n_1})\triangleq (x_{1,m_1},x_{1,m_1+1},\ldots,x_{1,m_1+k-1})$ and $\phi_2(m_2) = \phi_2(m_2; x_2^{n_2}) \triangleq (x_{2,m_2},x_{2,m_2+1},\ldots,x_{2,m_2+k-1})$, and attempts to correctly identify the \emph{phases} $(m_1,m_2)$. Similar to the point-to-point case, we define the \emph{rates} of the two sequences as
\[
R_1 \triangleq \frac{ \log_2{n_1}}{k} \quad \text{ and } \quad R_2 \triangleq \frac{\log_2 n_2}{k}.
\]
We note that every multiple access phase detection scheme induces two (multiset) \textit{codebooks} 
\begin{equation}
\label{eqn:c1}
 \mathcal{C}_1 = \{\phi_1(m_1)\suchthat m_1\in[n_1]\}\subseteq \Xc_1^k
\end{equation}
and 
\begin{equation}
\label{eqn:c2}
\mathcal{C}_2 = \{\phi_2(m_2)\suchthat m_2\in[n_2]\}\subseteq \Xc_2^k
\end{equation}
of rates $R_1$ and $R_2$ respectively. 
% Clearly for any phase detection scheme, the induced codebooks can be used as a MAC code for the same underlying channel. As we will see, the inverse direction is not true in general.

We discuss two different error criteria: the vanishing error criterion in Section~\ref{sec:mac-prob} and the zero error criterion in Section~\ref{sec:mac-advs}.

Under the vanishing error criterion, we assume that the phase pair $(M_1,M_2)$ is uniformly distributed over $[n_1] \times [n_2]$. The probability of error is defined as
\[
 P_e^{(k)} = \P\{(M_1,M_2) \neq (\mh_1(Y^k),\mh_2(Y^k))\}.
\]
A rate pair $(R_1,R_2)$ is said to be \emph{achievable} if, for a divergent sequence of $k$'s, there exist $(n_1,n_2,k)$ schemes with $\frac{\log{n_1}}{k} \ge R_1$, $\frac{\log{n_2}}{k} \ge R_2$, and $\lim_{k\to \infty}P_e^{(k)} = 0$. The \emph{vanishing error capacity region} $\Cr_\text{ve}$ is defined as the closure of the set of achievable rate pairs.

In Section~\ref{sec:mac-prob-limit}, we establish the vanishing error capacity region of multiple access phase detection. This region turns out to be \emph{strictly} included, in general, in the capacity region of its channel coding counterpart. This is in contrast to all models in the point-to-point case, in which phase detection either achieves the same best known rate or shares the same capacity as its channel coding counterpart. Due to the lack of synchronization between sequences, a phase detection scheme achieves at best the usual MAC capacity region without the time-sharing random variable. In Section~\ref{sec:mac-prob-code}, we provide a low-complexity ($O(k\log{k})$) sequence construction that achieves any rate pair in the capacity region.

Under the zero error criterion, a rate pair $(R_1,R_2)$ is said to be \emph{achievable} if, for a divergent sequence of $k$'s, there exist $(n_1,n_2,k)$ schemes with $\frac{\log{n_1}}{k} \ge R_1$ and $\frac{\log{n_2}}{k} \ge R_2$ such that $(m_1,m_2)$ can be recovered from $y^k$ with \emph{zero error} for any pair $(m_1,m_2)\in [n_1]\times [n_2]$. The \emph{zero error capacity region} $\Cr_\text{ze}$ is defined as the closure of the set of achievable rate pairs. We note that the problem of zero error phase detection in MACs is generally very difficult, as it is at least as hard as the zero error MAC coding problem, which in turn is open even in the simplest cases, e.g., the binary adder channel~\cite{Lindstrom1969,Kasami--Lin1978, Mattas--Ostergard2005, Urbanke--Li1998, Ordentlich--Shayevitz2016,Austrin--Kaski--Koivisto--Nederlof2016}. Nevertheless, in Section~\ref{sec:separation}, we demonstrate the distinction between the phase detection and the channel coding problems, by showing a separation between their capacity regions.

In Sections~\ref{sec:mac-advs-limit} and~\ref{sec:mac-advs-code}, we restrict our attention to a simple channel model, the modulo-2 addition channel with $\Xc_1 = \Xc_2 = \Yc = \{0,1\}$ and $Y = X_1 \oplus X_2$. For this channel, a rate pair $(R_1,R_2)$ is achievable if every element in the sumset 
\begin{equation}
\label{eqn:csum}
\Cc_\text{sum}\triangleq \{\phi_1(m_1)\oplus \phi_2(m_2)\suchthat m_1 \in [n_1], m_2 \in [n_2]\}
\end{equation}
can be uniquely expressed as an element in the induced codebook $\Cc_1$ plus an element in the induced codebook $\Cc_2$. Note that $\Cc_\text{sum}$ is defined as a regular set with distinct elements (rather than a multiset). Hence, any $\Cc_1$ and $\Cc_2$ induced by a valid scheme must also have distinct elements. 

Clearly, the zero-error channel coding capacity region $\{(R_1,R_2)\suchthat R_1 + R_2 \le 1\}$ is an outer bound for that of phase detection. In Section~\ref{sec:mac-advs-limit}, we establish the achievability of this region by a random construction that exploits properties of linear codes, in a way that resembles Wyner's linear Slepian--Wolf codes~\cite{Wyner1974}. We further provide in Section~\ref{sec:mac-advs-code} an explicit sequence construction that achieves this region, by exploiting
properties from finite field theory. As an consequence, the induced code from our phase detection sequences can be used for channel coding and achieve any rate pair in the zero-error capacity region, \emph{without using time sharing}\footnote{For other channel codes that achieve this region without using time sharing, see for example~\cite{Cohen--Litsyn--Vardy--Zemor1996, Ostergard--Kaikkonen1996}. For a channel code that achieves the rate pair $(1/2,1/2)$ with the \emph{same} codebook,
see~\cite{Lindstrom1969,Poltyrev--Snyders1995} for a construction utilizing the parity check matrix of a BCH code.}.

\section{Point-to-Point: Adversarial Noise}
In this section we discuss the adversarial noise model. We first examine whether the 
adversarial phase detection schemes achieve the best known rate for adversarial channel coding, namely the Gilbert--Varshamov (GV) bound~\cite{Gilbert1952,Varshamov1957}. 
\label{sec:advs}
\subsection{Fundamental Limit}
\label{sec:advs-limit}

\begin{theorem}
 \label{thm:advs-nkd-bound}
 An $(n,k)$ point-to-point phase detection scheme with minimum distance $d$ exists if
 \begin{equation}
 \label{eqn:advs-nkd-bound}
  n \le \frac{2^k}{16k \sum_{i=0}^d \binom{k}{i}}.
 \end{equation}
\end{theorem}

\begin{corollary}
\label{thm:advs-limit}
The capacity for adversarial phase detection is lower bounded by 
 \[
  C_\text{ad}(p) \geq  1-h(2p),
 \]
where $h(\cdot)$ is the binary entropy function.
\end{corollary}

We show the existence of a good sequence using the probabilistic method. We note that while several different proofs of the GV bound exist~\cite{Gilbert1952,Varshamov1957,Tolhuizen1997}, none of them seem to directly extend to our setting. This is simply due to the fact that there is a dependence between the codewords in the induced codebook. To alleviate this technical difficulty, we need the following well-known lemma.
\begin{lemma}[Lov\'{a}sz Local Lemma~\cite{Lovasz--Erdos1975}]
\label{lem:lll}
 Let $A_1,\ldots,A_N$ be a set of ``bad'' events with $\P(A_j) \le q < 1$, where each event 
$A_j$ is mutually independent of all but at most $L$ of the other events. If $4qL\le 1$, 
then
\[
 \P\left\{\cap_{j=1}^N A_j^c\right\}>0.
\]
\end{lemma}

\begin{IEEEproof}[Proof of Theorem~\ref{thm:advs-nkd-bound}]
  We generate the phase detection sequence $X^n$ i.i.d.$\sim \mathrm{Bern}(1/2)$ and apply minimum distance detection.  Let $\{A_j\}$ be the collection of events where the Hamming distance between a pair of codewords $\mathrm{wt}\bigl(\phi(m_1)\oplus \phi(m_2)\bigr) \le d$ where $m_1<m_2$. We have
\begin{align*}
 \P(A_j) &\stackrel{(a)}{=} \P\{\mathrm{wt}(Z^k) \le d, Z^k \text{ i.i.d.}\sim\mathrm{Bern}(1/2)\}\\
&= \sum_{i = 0}^{d} \binom{k}{i}\frac{1}{2^k}, %\\
%&\leq 2^{-k(1-h(2p))},
\end{align*}
where ($a$) follows since for any two distinct phases $m_1 \ne m_2$, the sum of the two codewords $\phi(m_1)\oplus \phi(m_2)$ is i.i.d.$\sim \mathrm{Bern}(1/2)$ even if they are overlapping subsequences of $X^n$. Now each $A_j$ is mutually independent of all other 
events, except for a set of at most $4kn$ events. This is because the random variable $\phi(m_1)\oplus \phi(m_2)$ is mutually independent of all $X_i$'s with $i \in [n]\setminus \{m_1-k+1,m_1-k+2,\ldots,m_1+k-1\} \setminus \{m_2-k+1,m_2-k+2,\ldots,m_2+k-1\}$, which excludes at most $4kn$ events. Applying Lemma~\ref{lem:lll}, the phase detection sequence $X^n$ 
has minimum distance greater than $d$ with positive probability
\[
 \P\left\{\cap_{j=1}^{\frac12 n(n-1)}A_j^c\right\}>0
\]
if
\begin{equation}
\label{eqn:nkd-2}
 16kn\sum_{i = 0}^{d} \binom{k}{i}\frac{1}{2^k} \le 1,
\end{equation}
or equivalently the condition in~\eqref{eqn:advs-nkd-bound}. This completes the proof of Theorem~\ref{thm:advs-nkd-bound}.
\end{IEEEproof}

\begin{IEEEproof}[Proof of Corollary~\ref{thm:advs-limit}]
Set $d = 2pk$ in~\eqref{eqn:advs-nkd-bound}. Applying the Hamming ball volume approximation
\[
\sum_{i = 0}^{2pk} \binom{k}{i}\leq 2^{kh(2p)}
\]
and plugging $R = \frac{\log{n}}{k}$ in~\eqref{eqn:nkd-2}, we have
\[
 R \le 1-h(2p) - \frac{\log (16k)}{k}.
\]
Letting $k\to \infty$, it follows that a rate $R$ is achievable if $R < 1-h(2p)$. 
% Applying the minimum distance decoder turns it into a good phase detection code.
\end{IEEEproof}

\begin{remark}
In the standard channel coding setup, a random codebook attains the GV bound with high probability. In contrast, the probability of randomly drawing a good scheme for our setup is exponentially small. This is most obvious in the noiseless case ($p=0$), where it is well known that the fraction occupied by de Bruijn sequences among all sequences vanishes exponentially fast~\cite{deBruijn1946}. 
\end{remark}

\subsection{Linear Phase Detection Schemes}
\label{sec:advs-code}
Theorem~\ref{thm:advs-nkd-bound} and Corollary~\ref{thm:advs-limit} showed the existence of a good adversarial phase detection scheme. Now, we discuss explicit constructions of such schemes. First, we ask whether phase detection schemes are ``equivalent'' to error-correcting codes in a certain sense. Clearly, any adversarial phase detection scheme induces a codebook that can be used as an error-correcting code for the corresponding adversarial channel coding problem. The converse direction seems more challenging. Given an error-correcting code, is it possible to ``chain up'' all or a sizable fraction of its codewords to create a sequence, and use the decoding rule as the detector? If so, what structure should such a code possess? In the following, we answer these questions for the class of \emph{linear} error-correcting codes. 

First, we note that in order to induce any error-correcting code with minimum distance $d>1$, the phase detection sequence $x^n$ should not contain $0^k$ as a contiguous subsequence, for otherwise a shift by one from that position would create a codeword that is at distance $1$ from $0^k$. Following that, an $(n,k)$ phase detection scheme is said to be \emph{linear} if $\mathcal{C}\cup \{0^k\}$, namely its induced codebook together with the zero codeword, forms a linear code. Let $r$ be the dimension of this linear code. Then, the length of the linear phase detection sequence is $n = 2^r-1$. 

\begin{theorem}
 \label{thm:lfsr}
A phase detection scheme with $n = 2^r-1$ is linear if and only if 
it is generated by a linear feedback shift register (LFSR) with a primitive characteristic 
polynomial  $a(z) = \sum_{i=0}^{r-1} a_i z^i + z^r$ over $GF(2)$, i.e.,
\begin{equation}
\label{eqn:lfsr}
 x_{r+j} = \sum_{i=0}^{r-1}a_i x_{i+j}, \quad j \in [n].
\end{equation}
 \end{theorem}
 
\begin{corollary}
\label{cor:linear}
 The non-zero codewords of a linear code of dimension $r$ can be 
chained up to a sequence of length $2^r-1$ if and only if any $r$ contiguous columns 
of the generator matrix
$
 G_{r\times k} = [\gv_1,\gv_2,\ldots,\gv_k]
$
are linearly independent, and 
\begin{equation}
\label{eqn:genMtx}
 \gv_{r+j} = \sum_{i=0}^{r-1}a_i \gv_{i+j},\quad j\in[k-r],
\end{equation}
where $a_i$'s are the coefficients of a primitive polynomial $a(z) = \sum_{i=0}^{r-1}a_i 
z^i + z^r$ over $GF(2)$.
\end{corollary}

\begin{IEEEproof}[Proof of Theorem~\ref{thm:lfsr}]
  To prove sufficiency, suppose that $x^n$ is generated by an LFSR with a primitive characteristic polynomial in~\eqref{eqn:lfsr} and a nonzero initial state vector $(x_1,x_2,\dots,x_r)$. Then, every length-$r$ string except $0^r$ 
  occurs exactly once in $x^n$ (see~\cite[Theorem 8.33]{Lidl--Niederreiter2008}). It follows that for any distinct codewords $\phi(m_1) = 
c^k$ and $\phi(m_2) = d^k$, there exists $\phi(m_3) = e^k$ such that $c_j + d_j = 
e_j$ for $j \in [r]$. For $r < j \le k$, $c_{j} + d_{j} = e_{j}$ follows since the 
sequence is generated by an LFSR of degree $r$.
% \begin{align*}
% c_{r+j} + d_{r+j} &= \sum_{i=0}^{r-1}a_i c_{i+j} + \sum_{i=0}^{r-1}a_id_{i+j}\\
% &= \sum_{i=0}^{r-1}a_i (c_{i+j}+ d_{i+j}).
% \end{align*}

For necessity, let $x^n$ be a sequence associated with a linear phase detection scheme. We show that the first 
$r$ columns $\gv_1,\ldots,\gv_r$ of the generator matrix $G_{r\times k}= [\gv_1,\gv_2,\ldots,\gv_k]$ must be linearly independent. 
Assuming that contrary, there exist $f_1,\ldots,f_r \in \{0,1\}$ not all zero such that
\begin{equation}
\label{eqn:rank}
 \sum_{i=1}^r f_i \gv_i = {\bf 0}. 
\end{equation}
Let $[x_1,\ldots,x_k] = [u_1,\ldots,u_r]G_{r\times k}$. Multiplying 
both sides of~\eqref{eqn:rank} by  $[u_1,\ldots,u_r]$, we have $\sum_{i=1}^r f_i x_i = 0$. Applying this to
every codeword in $\mathcal{C}$, and recalling that the codewords are all contiguous subsequences of $x^n$, we have
\[
 \sum_{i=1}^r f_i x_{i+j} = 0, \quad j\in[n].
\]
Let $i_0 = \max\{i\in[r]\suchthat f_i=1\}$. If $i_0 = 1$, then $x^n$ has to be $0^n$, 
in contradiction. For $i_0 > 1$, we have
\[
 x_{j+i_0} = \sum_{i=1}^{i_0-1} f_i x_{i+j},\quad j\in [n],
\]
which implies $x^n$ is generated by an LFSR of degree $i_0-1 < r$. But this 
contradicts the fact that $x^n$ is of length $2^r-1$ and all codewords $\phi(m), m\in[n]$, 
are distinct. 

Now, since the first $r$ columns of $G_{r\times k}$ are linearly independent, there 
exist $a_0, \ldots, a_{r-1}$ such that 
\[
 \gv_{r+1} = \sum_{i=0}^{r-1} a_i \gv_{i+1}.
\]
From this it follows that~\eqref{eqn:lfsr} holds and $x^n$ is generated by an LFSR. 
Finally, an LFSR sequence is of maximum length if and only if the characteristic polynomial 
is primitive.
\end{IEEEproof}

\begin{IEEEproof}[Proof of Corollary~\ref{cor:linear}]
 The sufficiency follows since for a linear code, the relation~\eqref{eqn:genMtx} implies~\eqref{eqn:lfsr}. The necessity follows the same way as the necessity in Theorem~\ref{thm:lfsr}.
\end{IEEEproof}

\begin{remark}
As an application of Theorem~\ref{thm:lfsr}, we can design a card trick for adversarial crowds. Picking the primitive polynomial $a(z) = z^5 + z^4 + z^2 + z + 1$ and $k = 9$, we get a sequence of length $n =31$ and minimum distance $d =3$. Ordering cards according to this sequence, the magician can now correct one lie out of 9 contiguous color reads.
\end{remark}

\begin{remark}
When the characteristic polynomial of the LFSR is irreducible but not primitive, the sequence it generates has length $t$, which equals the order of the characteristic polynomial. Depending on the initial state $x^r$, the LFSR generates one out of $s=\frac{2^r-1}{t}$ \emph{disjoint} sequences $x^t{(1)}, \ldots, x^t{(s)}$. The length-$k$ contiguous subsequences of each sequence $\mathcal{C}^{(i)} = \{x_{m}^{m+k-1}(i)\suchthat m \in [t]\}$ together with the zero codeword form a linear code
$
 \cup_{i=1}^s\mathcal{C}^{(i)} \cup \{0^k\}.
$
Conversely for a linear code, if the first $r$ columns of its generator matrix are linearly independent and~\eqref{eqn:genMtx} holds with $a_i$'s being the coefficients of an irreducible but not primitive polynomial of order $t$, then its nonzero codewords can be partitioned into $s$ equal size subsets, each of which can be chained up to a phase detection sequence.
\end{remark}

We now provide two results on the performance of linear phase detection schemes. In Theorem~\ref{thm:concentration}, we cite a known result from~\cite[Theorem 8.85]{Lidl--Niederreiter2008} on asymptotic relative distance, which improves upon~\cite[Theorem~1]{Kumar--Wei1992}. Then, inspired by a linear programing bound for LDPC codes~\cite{Ben-Haim--Litsyn2006}, we provide in Theorem~\ref{thm:ldpc} an upper bound on the sequence length of a linear phase detection scheme of a given minimum distance, using the linear programing method originated by Delsarte~\cite{Delsarte1973}. 

\begin{theorem}[Theorem 8.85~\cite{Lidl--Niederreiter2008}]
\label{thm:concentration}
 For every $(n,k)$ linear phase detection scheme, for every $m \in [n]$,
 \[
  \left|\emph{wt}(x_m^{m+k-1})-\frac{k}{2}\right| \le \sqrt{n} \left(\frac{\log{n}}{\pi}+1\right).
 \]
In particular, for $(n,k)$ such that $\lim_{k\to \infty} \frac{\sqrt{n}\log{n}}{k}= 0$, the relative distance of the induced code converges to 
\begin{equation}
\label{eqn:relative-d}
 \lim_{k \to \infty} \frac{\;d\;}{k} = \frac 12.
\end{equation}

\end{theorem}

\begin{remark}
 We note a similar result in~\cite[Theorem~1]{Kumar--Wei1992}, which claims~\eqref{eqn:relative-d} for every $0 < \mu \le 1$ and $k = \mu n$. Theorem~\ref{thm:concentration} improves upon~\cite{Kumar--Wei1992} by allowing $k$ to be sublinear in $n$.
\end{remark}

% Before presenting the next result, let us recall some definitions. For $t \in [k], z \in \mathbb{R}$, the Krawtchouk polynomial~\cite[Ch.~5. \S~2]{MacWilliams--Sloane1977a}~\cite{MacWilliams--Sloane1977b} is defined as

For the next result, we need the following definitions. For $t \in [k]$ and $z \in \mathbb{R}$, let
\[
 K_t(z) = \sum_{j=0}^t (-1)^j \binom{z}{j} \binom{l-z}{t-j}
\]
be the Krawtchouk polynomial~\cite[Ch.~5. \S~2]{MacWilliams--Sloane1977a}~\cite{MacWilliams--Sloane1977b}, where the binomial coefficient for $z \in \mathbb{R}$ is defined as
$
 \binom{z}{i} = \frac{z(z-1)(z-2)\cdots (z-i+1)}{i\, !}.
$
For large $k$, the exponent of $K_t(z)$ can be approximated as~\cite[Equation (40)]{Ben-Haim--Litsyn2006} 
\[
\frac{1}{k}\log{K_{\lfloor pk\rfloor}(\lfloor\lambda k \rfloor)} = h(p) + \text{Int}(p,\lambda) + o(1),
\]
where 
\begin{align}
  &\text{Int}(p,\lambda) \nonumber\\
  &= \int_0^{\lambda} \log\left(\frac{1-2p + \sqrt{(1-2p)^2 - 4(1-y)y}}{2(1-y)}\right)dy. \label{eqn:int}
\end{align}

\begin{theorem}
 \label{thm:ldpc}
 Every $(n,k)$ linear phase detection scheme with length $n =2^r-1$ and minimum distance $d = 2t + 1$ must satisfy 
 \[
  2^r \cdot \frac{K_{t}^2(ic)\binom{(k-r)/c^2}{i}c^{2i}}{\binom{k}{t}} \le 2^k
 \]
 for every $i\in [k]$ such that $ic^2 \le k - r$. Here $c$ is the number of nonzero coefficients of the characteristic polynomial $a(z) = \sum_{j=0}^r a_j z^j$.
\end{theorem}

\begin{remark}
Compared to Delsarte's linear programing bound for channel codes~\cite{Delsarte1973},
% %  \[
% %   2^r \cdot \sum_{i=0}^t \binom{k}{i} \le 2^k,
% %  \]
the bound in Theorem~\ref{thm:ldpc} can sometimes be better. For example, when $r = 20,t = 5$, and $c = 3$, the linear programing bound yields $k \ge 41$, while Theorem~\ref{thm:ldpc} requires $k \ge 42$. We note, however, that with further optimization for these specific parameters, the best known channel coding upper bound is $k \ge 43$~\cite{Grassl2007}.
\end{remark}

\begin{remark}
 For low-complexity LFSR implementation, it may be desirable to choose a characteristic polynomial with low coefficient weight. According to a conjecture in finite field theory~\cite{Mullen--Shparlinski1996, Golomb2007}, there are infinitely many primitive polynomials with coefficient weight $c = 3$. For this class of primitive polynomials, Theorem~\ref{thm:ldpc} implies that when the adversarial channel can flip at most a fraction $p$ of the inputs, the rate of the linear phase detection scheme must satisfy
 \begin{align}
  &\max_{0\le \mu \le \frac{1-R}{9}}\left\{2\mu \log{3} +  h\left(\tfrac{9\mu}{1-R}\right) \tfrac{(1-R)}{9} +2\, \text{Int}(p,3\mu) \right\} \nonumber\\
  &\le 1 - h(p) - R,  \label{eqn:new-ub}
 \end{align}
where $\text{Int}(p,3\mu)$ is given in~\eqref{eqn:int}.
This bound can sometimes be better than the second MRRW bound~\cite{McEliece--Rodemich--Rumsey--Welch1977}, which is the best known asymptotic upper bound for binary channel codes. For example, when $p = 0.05$, the second MRRW bound requires $R \le 0.6927$. However, $p=0.05$ and $R = 0.6927$ violate condition~\eqref{eqn:new-ub} when $\mu = 0.03073$.
\end{remark}

\begin{IEEEproof}[Proof of Theorem~\ref{thm:ldpc}]
 Following the same line of reasoning as in Section II-C (29)--(36) and (48)--(49) of~\cite{Ben-Haim--Litsyn2006}, we have for every $\a\in [k]$,
\begin{equation}
 \label{eqn:krawtchouk}
  2^r \cdot \frac{K_{t}^2(\a) B_{\a}}{\binom{k}{t}} \le 2^k,
\end{equation}
where $B_{\a}$ is the number of codewords of weight $\a$ in the dual code of the  linear code induced by the phase detection scheme. Now we show that when the coefficient weight of the characteristic polynomial is $c$, for every $i \in [k]$ such that $ic^2 \le k-r$, we can lower bound
\begin{equation}
\label{eqn:weight}
 B_{ic} \ge \binom{(k-r)/c^2}{i} c^{2i}.
\end{equation}

To that end, note that our $(k-r)\times k$ parity check matrix, which is also the generator matrix of the dual code, can be written in the following form
$$
% H = 
% \left[ 
%   \begin{array}{cccccccccc}
%     1 & a_1 & a_2 & \cdots & a_{r-1} & 1 & 0 & 0 &\cdots & 0\\
%     0 & 1 & a_1 & a_2 & \cdots & a_{r-1} & 1 & 0 & \cdots & 0\\
%     \vdots&\vdots&\vdots&\vdots&\vdots&\vdots&\vdots&\vdots&\vdots&\vdots\\
%     0 & 0 &\cdots &0 & 1 & a_1 & a_2 & \cdots & a_{r-1} & 1 \\
%   \end{array}  
% \right]_{(k-r)\times k.}
\left[ 
  \begin{array}{ccccccccc}
    1 & a_1 & \cdots & a_{r-1} & 1 & 0 & 0 &\cdots & 0\\
    0 & 1 & a_1 & \cdots & a_{r-1} & 1 & 0 & \cdots & 0\\
    \vdots&\vdots&\vdots&\vdots&\vdots&\vdots&\vdots&\vdots&\vdots\\
    0 & 0 &\cdots &0 & 1 & a_1 & \cdots & a_{r-1} & 1 \\
  \end{array}  
\right]_{.}
$$
A weight $ic$ codeword of the dual code could come from the sum of $i$ rows of $H$ whose nonzero elements (the $1$'s) are in disjoint columns. We lower bound the number of such codewords. First, we select an arbitrary row from the $(k-r)$ rows. Since each row of $H$ has weight $c$, the locations of the $1$'s in the chosen row overlap that of at most $c^2$ rows (including itself). Then a second row is chosen from the $(k-r-c^2)$ remaining non-overlapping rows. We continue in this manner until we obtain $i$ rows (we will not exhaust all rows provided that $ic^2 \le k-r$). Hence, the number of choices is lower bounded by
\begin{align*}
 &\frac{1}{i\,!}(k-r)(k-r-c^2)\cdots(k-r-(i-1)c^2)\nonumber\\
 &={(k-r)/c^2\choose i}c^{2i},
\end{align*}
which establishes~\eqref{eqn:weight}. Plugging~\eqref{eqn:weight} into~\eqref{eqn:krawtchouk} with $\a = ic$ completes the proof. 
\end{IEEEproof}

\section{Point-to-Point: Probabilistic Noise,\\ Vanishing Error}
\label{sec:prob}
In this section we discuss the probabilistic noise model with a vanishing error criterion. We first show that the capacity in this case coincides with the Shannon capacity of the observation channel. We then proceed to describe a low-complexity  coding construction, based on a concatenation of a channel code and a de Bruijn sequence, that approaches this fundamental limit. 
\subsection{Fundamental Limit}
\label{sec:prob-limit}

\begin{theorem}
\label{thm:prob-limit}
 The vanishing error capacity for probabilistic phase detection over a channel $p(y|x)$ is
 \[
  C_\text{ve} = \max_{p(x)}I(X;Y).
 \]
\end{theorem}

Before we proceed to the proof, we need a technical lemma. We denote the typical set of length-$k$ vectors corresponding to $(X,Y)$ by
\begin{align*}
&\mathcal{T}_\e^{(k)}(X,Y)\\
&:= \Big\{(x^k,y^k)\suchthat \left|\frac{\#\{i\suchthat (x_i,y_i) = (x,y)\}}{n} - p(x,y)\right|\\
&\hspace{2em}\le \e p(x,y)\text{ for all } x \in \Xc, y \in \Yc\Big\}.
\end{align*}
 
\begin{lemma}[Lemma 24.2~\cite{El-Gamal--Kim2011}]
\label{lem:iid}
Let $(X,Y)\sim p(x,y) \neq p(x)p(y)$ and $(X^n,Y^n)\sim \prod_{i=1}^n p_{X,Y}(x_i,y_i)$. 
If $\e > 0$ is sufficiently small, then there exists $\gamma (\e)>0$ that depends only on 
$p(x,y)$ such that
\begin{equation}
\label{eqn:overlap}
 \P\{(X_{m}^{m+k-1},Y^k)\in \Tc_\e^{(k)}(X,Y) \} \le 2^{-k \gamma(\e)}
\end{equation}
for every $m>1$. Moreover, for non-overlapping sequences, i.e., for $k+1 \le m \le n-k 
+1$, \begin{equation}
\label{eqn:indep}
 \P\{(X_{m}^{m+k-1},Y^k)\in \Tc_\e^{(k)}(X,Y)\} \le 2^{-k(I(X;Y)-\delta(\e))},
\end{equation}
where $\delta(\e)$ tends to zero as $\e \to 0$.
\end{lemma}

\begin{IEEEproof}[Proof of Theorem~\ref{thm:prob-limit}]
Clearly, any phase detection sequence is also a channel code. Thus, the above rate cannot be exceeded. We proceed to prove the achievability. Recall $\phi(m) = x_{m}^{m+k-1}$. 

\smallskip
{\it Phase detection sequence generation.} We generate the sequence $X^n$ i.i.d.$\sim 
p(x)$.

\smallskip
{\it Detection.} Upon receiving $y^k$, the detector declares $\mh$ is the phase estimate if 
it is the unique phase such that $(\phi(\mh),y^k) \in \Tc_\e^{(k)}(X,Y)$; otherwise---if there is 
none or more than one---it declares an error. 

\smallskip
{\it Analysis of the probability of error.} Without loss of generality, we assume the phase $M=1$. The detector makes 
an error only if one or more of the following events occurs:
\begin{align*}
 \mathcal{E}_1 &= \{(\phi(1),Y^k) \notin \Tc_\e^{(k)}(X,Y)\},\\
 \mathcal{E}_2 &= \{(\phi(m),Y^k) \in \Tc_\e^{(k)}(X,Y) \text{ for some } m \ne 1\}.
\end{align*}
By the law of large numbers, $\P(\mathcal{E}_1)$ tends to zero as $k \to \infty$. For the 
second term, we have
\begin{align*}
 &\P(\mathcal{E}_2) \\
 &\le \left(\sum_{m=2}^k + \sum_{m=n-k+2}^n\right) 
\P\{(\phi(m),Y^k) \in \Tc_\e^{(k)}(X,Y)\} \\
&\hspace{2em} + \sum_{m=k+1}^{n-k+1} \P\{(\phi(m),Y^k) \in \Tc_\e^{(k)}(X,Y)\}\\
% &\stackrel{(a)}{\le} 2(k-1)2^{-k\gamma(\e)} + (n-2k+1)2^{-k(I(X;Y)-\delta(\e))}\\
&\stackrel{(a)}{=} 2(k-1)2^{-k\gamma(\e)} + (2^{kR}-2k+1)2^{-k(I(X;Y)-\delta(\e))},
\end{align*}
which tends to zero as $k \to \infty$ if $R < I(X;Y)-\delta(\e)$. Here the first and the
second terms in ($a$) follow from~\eqref{eqn:overlap}~and~\eqref{eqn:indep} respectively.
Letting $\e \to 0$ completes the proof. 
\end{IEEEproof}

\begin{example}\label{ex:GPS}
Consider the case of GPS signaling. For GPS, the binary (BPSK) symbol duration is about 1$\mu$sec, and the length of the underlying Gold code sequence is $n=1023$. Consider a typical observation time of $1$ second, which corresponds to a repetition factor $N\approx 1000$ and $k\approx 1e6$ binary observations. A correlator receiver can thus increase the SNR by about $60\textrm{dB}$ by coherently integrating over this sequence (assuming symbol timing has been recovered). Due to the good autocorrelation structure of the Gold code, an SNR of $30\textrm{dB}$ is typically sufficient in order to distinguish the correct phase (out of the $1023$ possibilities, and typically also over several Doppler hypotheses), with a small enough error probability. Namely, one can operate at an SNR of $-30\textrm{dB}$, and provide positioning with uncertainty of $1023\,\mu$sec; multiplied by the speed of light, this yields a positioning modulo $\approx 30,000$ km, which is sufficient as it is of the same order of the distance to the satellites. 

Let us now show that one can significantly improve sensitivity using a more general phase detection sequence. Using the same observation period of $1$ second, let us assume a much lower SNR of $-44\textrm{dB}$. Using the Gaussian capacity formula and Theorem~\ref{thm:prob-limit}, we have that 
\begin{align*}
\frac{\log{n}}{k} \approx \frac{1}{2}\log_2(1+SNR)  
\end{align*}
can be asymptotically achieved. Using our $k=1e6$ and solving for $n$, we get that the largest $n$ that can be supported is $n\approx 4e8$. Since this large $n$ is also (much) larger than $k$, we can in principle design a phase detection sequence with roughly these parameters that attains a low error probability. This will reliably find our distance to the satellite with an uncertainty of about $120$ billion km, a huge overkill, but saves $14\textrm{dB}$ in the SNR relative to the competing GPS solution operating with the same observation time. To make the comparison more precise, one should look more carefully at many important details such as the exact error probability performance, the effect of multiple Doppler hypotheses, complexity of detection, and accounting for multiple satellites. Most of these issues are beyond the scope of this paper. In the next subsection and in Secion~\ref{sec:mac-advs} we discuss the issues of complexity and multiple sequences. 
\end{example}

\subsection{A Low-Complexity Construction}
\label{sec:prob-code}
Now we present a sequence construction with low-complexity detection that 
achieves the capacity asymptotically. The construction consists of three main ingredients:
\begin{enumerate}
\item a de Bruijn sequence with an efficient decoding algorithm~\cite{Tuliani2001},
%  \item a linear feedback shift register with a primitive polynomial as its 
% connection function to generate a \emph{maximal length sequence} (almost the same as the 
% de Bruijn sequence except that it the all zero string does not occur as a subsequence),

\item a capacity achieving low-complexity code, e.g. a polar code~\cite{Arikan2009}, that protects the de Bruijn 
sequence against noise, and

\item an i.i.d. synchronization sequence, which is known at the detector {\it a priori}, that 
allows the detector to find the block boundary.
\end{enumerate}
The details are as follows.

\begin{figure*}[t!]
\centering
% \hspace{-1.5em}
 \def\svgscale{2.1}
 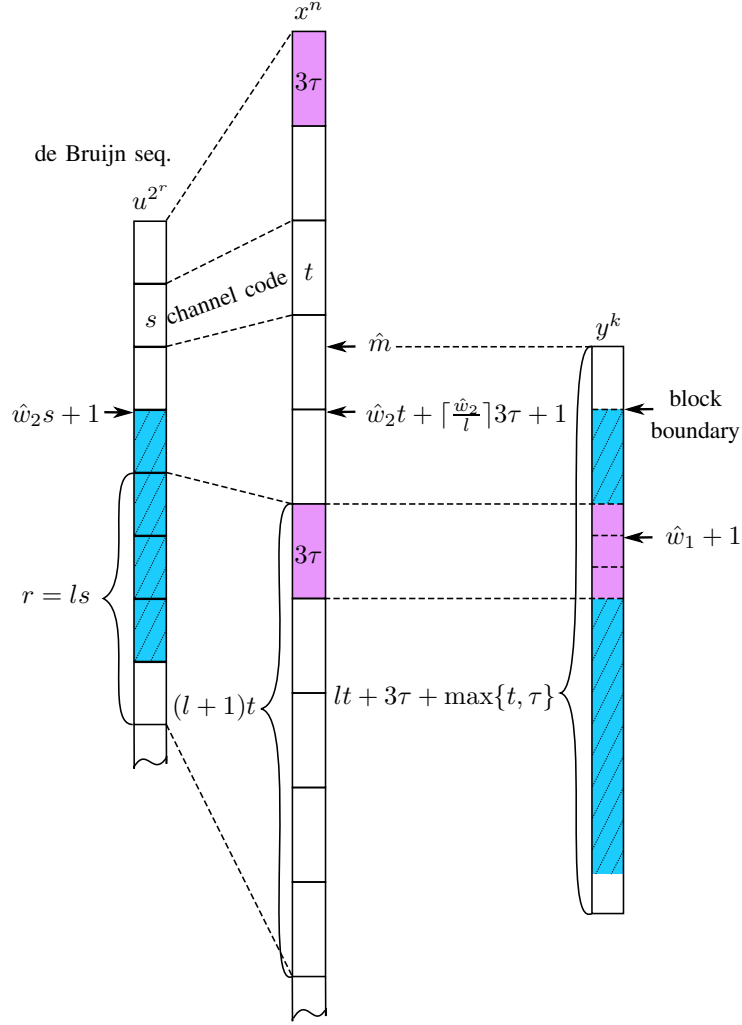
 \caption{Construction for probabilistic phase detection.}
 \label{fig:prob-code}
\end{figure*}

\smallskip
{\it Phase detection sequence design.} 
\iffalse
Let $a(z) = \sum_{i=0}^{r-1} a_i z^i + 
z^{r}$ be a primitive polynomial over $GF(2)$. A sequence of length $2^{r}-1$ 
sequence is generated through the linear feedback shift register with connection function 
$a(z)$, i.e., $$u_{j+r} = \sum_{i=0}^{r-1}a_i u_{j+i}, \;\;j = 
1,2,\ldots,2^r-r-1,$$ 
with the initial $u^{r}$ being not all zeros (for example, $u_1 = 1, u_2 = 
\cdots = u_{r} = 0$). This generates a maximal length circulant sequence with the 
property that any $r$ contiguous bits uniquely determines its location in the sequence.
\fi
We design a de Bruijn sequence $u^{2^r}$ of order $r$ according the method in~\cite{Tuliani2001}. To encode it to a phase detection sequence $x^n$, we let $r = sl$, where $s$ and $l$ are integers. The de Bruijn sequence is chopped up into length-$s$ chunks, each of which is encoded into a length-$t$ codeword using a channel code of rate $R_\mathrm{ch} = s/t$. Then, a synchronization sequence $b^{3\tau}$ is generated i.i.d.$\sim p(x)$, where the parameter $\tau$ is a linear function of $t$, i.e., $\tau = c_1 t + c_2$ for some constants $c_1>0$ and $c_2$. Below we use $\tau = t$, but $\tau \neq t$ will prove useful later in Section~\ref{sec:mac-prob-code}. This sequence $b^{3\tau}$ is inserted every $l$ blocks. The middle chunk of the synchronization sequence $b_{\tau+1}^{2\tau}$ is given to the detector. The chunks $b^{\tau}$ and $b_{2\tau+1}^{3\tau}$ play the role of ``guarding bits'' between codewords and the middle chunk $b_{\tau+1}^{2\tau}$. Their purpose is to simplify the analysis of the error probability event associated with the synchronization detection (later denoted $\Ec_1$), as will become clear in the sequel. This is illustrated in Figure~\ref{fig:prob-code}. 

\smallskip
{\it Detection.} We choose the length of the detection window to be
\begin{equation}
\label{eqn:dec-window}
k = lt+3\tau+\max\{t,\tau\}.
\end{equation}
The extra $\max\{t,\tau\}$ symbols are the margin to ensure there are $l$ complete channel code blocks and a complete synchronization sequence in the received sequence. Upon receiving $y^k$, the detector first finds an $\wh_1 \in \{0\} \cup [k-\tau]$ such that $(b_{\tau+1}^{2\tau},y_{\wh_1 +1}^{\wh_1+\tau}) \in \mathcal{T}_\e^{(\tau)}(X,Y)$. If there are more than one, it chooses the smallest index. It declares an error if there is none. This determines the block boundary of the channel code blocks, i.e., a complete block starts from index $(\wh_1-\tau \mod t) + 1$ in $y^k$. By design, there are at least $l$ complete channel code blocks in $y^k$ (the dashed-line parts in $y^k$ of Figure~\ref{fig:prob-code}). The detector then applies the channel decoder to recover $l$ blocks of messages. This corresponds to $ls = r$ contiguous bits in the de Bruijn sequence (the dashed-line parts in the $u$ sequence of Figure~\ref{fig:prob-code}), which uniquely determines the location of these bits $u_{s\wh_2 + 1}^{s\wh_2+r}$ via the de Bruijn decoder of~\cite{Tuliani2001}. The phase estimate is then declared as
\[
 \mh = \wh_2 t + \left\lceil\frac{\wh_2}{l}\right\rceil 3\tau + 1 - (\wh_1 -\tau \mod 
t).
\]

\smallskip
{\it Analysis of the probability of error.} For clarity of notation, we set $\tau = t$ in the following analysis. Similar analysis can be done for other linear functions of $t$.  %We generate the synchronization sequence $B^{3t}$ i.i.d.$\sim p(x)$. 
% It is known that by randomizing the frozen bits, the polar coding ensembles are i.i.d.$\sim 
% p(x)$~\cite{Arikan2009,Honda--Yamamoto2013,Sasoglu--Wang2013}. 
Let $W_1$ be the actual index of the noisy version of $B_{\tau}$ in $Y^k$. The detector makes an error only if at least one of the following events occurs:
\begin{align*}
 \mathcal{E}_1 &= \{W_1 \ne \Wh_1\},\\
 \mathcal{E}_2 &= \{\text{an error in channel decoding}\}.
\end{align*}
Given $\mathcal{E}_1^c \cap \mathcal{E}_2^c$, the de Bruijn decoder can figure out the phase of the decoded $r$ bits with zero error. Since we are using a good channel code, we have $\P(\mathcal{E}_2 \cap \mathcal{E}_1^c)\to 0$ as $t\to\infty$. To bound $\P(\mathcal{E}_1)$, assume for convenience and without loss of generality that $W_1=t-1$.  We have
{\allowdisplaybreaks
\begin{align*}
 \P(\mathcal{E}_1)&= \P\{(B_{t+1}^{2t},Y_{w_1+1}^{w_1+t}) \in \mathcal{T}_\e^{(t)} \text{ for some } w_1 \ne t-1\}\\
&\le \sum_{w_1 = 0}^{t-2} \P\{(B_{t+1}^{2t},Y_{w_1+1}^{w_1+t}) \in \mathcal{T}_\e^{(t)}\} \\
&\hspace{2em} + \sum_{w_1 = t}^{2t-2} \P\{(B_{t+1}^{2t},Y_{w_1+1}^{w_1+t}) \in \mathcal{T}_\e^{(t)}\}\\
&\hspace{2em} + \sum_{w_1 = 2t-1}^{k-t} \P\{(B_{t+1}^{2t},Y_{w_1+1}^{w_1+t}) 
\in \mathcal{T}_\e^{(t)}\}\\
&\stackrel{(a)}{\le} (t-1) 2^{-t \gamma(\e)} + (t-1)2^{-t \gamma(\e)}\\
&\hspace{2em}+ ((l+1)t+2)2^{-t(I(X;Y)-\delta(\e))},
\end{align*}}%
which, for fixed $l$, tends to zero as $t \to \infty$.
Here, the first term in ($a$) follows from Lemma~\ref{lem:iid} and the fact that $B_{t+1}^{2t}$ 
and its preceding guarding block $B^{t}$ are i.i.d.$\sim p(x)$. The second term follows since $B_{t+1}^{2t}$
and its succeeding guarding block $B_{2t+1}^{3t}$ are i.i.d.$\sim p(x)$. The third term follows by virtue of the packing lemma~\cite[Lemma~3.1]{El-Gamal--Kim2011}, since any length-$t$ chunk from two channel code blocks is independent of $B_{t+1}^{2t}$. Note the role of the ``guarding bits'' here is to make sure $Y_{w_1+1}^{w_1+t}$ never overlaps with both $B_{t+1}^{2t}$ and a codeword, as we cannot generally assume too much about the statistics of a specific codeword. Therefore, the probability of error averaged over all possible realizations of $B^{3t}$ tends to zero as $t \to \infty$. It follows that a good deterministic sequence $b^{3t}$ exists (in fact, most choices are good).

\smallskip
{\it Rate.} 
By design, the rate of the sequence is
\begin{align}
 R &= \frac{\log n}{k}\nonumber\\
 &= \frac{\log\left[2^{r}\frac{lt+3\tau}{ls}\right]}{lt+3\tau+\max\{t,\tau\}} \label{eqn:p2p-rate}\\
 &\stackrel{(a)}{=} \frac{\log\left[2^{r}\frac{(l+3)t}{ls}\right]}{(l+4)t} \nonumber\\
 &= R_\mathrm{code}\left(1-\frac{4}{l+4}\right) \nonumber\\
 &+ \frac{R_\mathrm{code}}{s(l+4)}\log\left[\frac{1}{R_\mathrm{code}}\left(1+\frac 
3l\right)\right], \nonumber
\end{align}
which, for fixed $R_\mathrm{code}$ and $l$, tends to $R_\mathrm{code}\left(1-\frac{4}{l+4}\right)$ as $s \to \infty$.  Choosing a large $l$ and a capacity achieving code for the underlying channel $p(y|x)$ ensures the rate of the 
phase detection sequence can be as close to $C_\mathrm{prob}$ as desired. Note that in step ($a$), we set $\tau = t$. But one can verify that the rate approaches capacity $C_\mathrm{prob}$ for other choices of $\tau$. 

\smallskip
{\it Complexity.} Finding the block boundary is $O(k)$ complexity. Recalling that $r$ is linear in $k$ and using the method 
of~\cite{Tuliani2001}, decoding the de Bruijn sequence is $O(k\log k)$ complexity. There exist capacity achieving channel codes  
with $O(k \log k)$ decoding complexity, e.g., polar codes~\cite{Arikan2009}. 
Therefore, the overall detection complexity is $O(k \log k)$.

\begin{remark}%[$(R_\text{pc},l,t,\tau)$ Point-to-Point Phase Detection Sequence]
 \label{rmk:p2p-code}
 For future reference, we refer to the above construction an \emph{$(R_\text{ch},l,t,\tau)$ point-to-point phase detection sequence}. Once these four parameters are given, $s = tR_\text{ch}, r = ls$, and both $k$ and $R$ can be expressed as in~\eqref{eqn:dec-window} and~\eqref{eqn:p2p-rate}. As shown above, an $(R_\text{ch},l,t,\tau)$ point-to-point phase detection sequence has detection complexity $O(k \log{k})$. Moreover, for $\tau = c_1t + c_2$ with some constants $c_1 > 0$ and $c_2$, the achievable rate of the sequence satisfies
 \[
  \lim_{l\to \infty}\lim_{t \to \infty} R(R_\text{ch},l,t,\tau) = R_\text{ch}.
  \]
 This construction will also prove useful in Section~\ref{sec:mac-prob-code}. 
\end{remark}

\begin{remark}
\label{rmk:eqvt-avg}
It appears plausible that the synchronization sequence could  be discarded, and that the codeword boundary could be determined as part of the detection process. This coding scheme, in a sense, shows the equivalence between error-correcting codes and phase detection schemes for the probabilistic setting.
\end{remark}

\begin{remark}
\label{rmk:eqvt-max}
 Our analysis for the point-to-point phase detection problem in the probabilistic noise model assumed a uniformly distributed phase, which in channel coding terms corresponds to an average error probability criterion. In channel coding, the capacity under a more stringent maximal error probability criterion remains the same; this is easily shown by throwing away the worse half of a good average error probability codebook. In the sequence phase detection problem however, it is not immediately clear whether the capacity remains the same, as throwing bad codewords can significantly shorten the sequence. However, using our specific construction above and using a maximal error capacity achieving channel code (which may increase the detection complexity), we can show that the resulting phase detection sequence is capacity achieving under maximal error probability criterion. 
\end{remark}

\section{Point-to-Point: Probabilistic Noise,\\ Zero Error}
\label{sec:prob-zero}

In this section, we consider zero error phase detection. Let $\a(G)$ denote the \emph{independence number} of a graph $G$, i.e., the cardinality of a maximum independent set of $G$.
% Let us recall some notation on a graph $G = (\Xc,E)$. A subset of the vertex set $\Xc$ is called an \emph{independent set} if no two vertices in the set are connected by an edge. Let $\a(G)$ denote the \emph{independence number} of $G$, i.e., the cardinality of the largest independent set. 
Then, the \emph{Shannon capacity of a graph} $G$ can be defined as~\cite{Shannon1956}
\[
C(G) \triangleq \sup_{k} \frac{\log{\a(G^k)}}{k} = \lim_{k \to \infty} \frac{\log{\a(G^k)}}{k},
\]
where $G^k$ is the $k$-fold strong product of $G$ (see definition in Section~\ref{sec:intro-p2p}). It is well known that $C(G)$ is the zero error capacity of any channel $p(y|x)$ with confusion graph $G$. An explicit expression for $C(G)$ is unknown. Nevertheless, the following theorem shows that $C(G)$ is also the fundamental limit in the zero error phase detection setting. 

\begin{theorem}
 \label{thm:prob-zero-limit}
 The zero error capacity for phase detection in a channel with confusion graph $G$ coincides with the Shannon capacity of this graph, i.e.,
 \[
  C_\text{ze}(G) = C(G).
 \]
\end{theorem}

\begin{IEEEproof}
Again, the induced codebook of every phase detection scheme is also a good channel code for the same confusion graph, and thus $C_\text{ze}(G) \le C(G)$. For the other direction, we show that every channel code of rate $R$ can be used to construct a phase detection scheme with the same rate in the asymptotic limit. 

To this end, we first note that the rate $\log{\a(G)}$ can be readily achieved. This can be done by employing a one-shot zero error channel code of the same rate (which exists by definition), and using it to construct a de Bruijn sequence of alphabet size $\a(G)$ and order $k$ (cf. the existence of de Buijn sequences of any alphabet size and any order~\cite{Martin1934}). When $C(G) > \log{\a(G)}$, which means that the graph capacity can be achieved only by block coding over the product graph, then concatenating a length-$k$ zero error channel code according to a de Bruijn sequence of alphabet size $\a(G^k)$ (a naive extension of the one-shot approach above) does not immediately work (see also Remark \ref{rmk:last-one}). This is because the phase detector may not always know where a complete codeword starts or ends, which may result in detection errors. In what follows, we design a novel zero error synchronization sequence that enables the detector to  determine the block boundary without error, and with vanishing loss in rate.

% Our construction consists of two main ingredients:
% \begin{enumerate}
%  \item the concatenation of a de Bruijn sequence over extended alphabet with a zero error channel code, in a way that is similar to the construction in Section~\ref{sec:prob-code}, and
%  \item a novel design for the zero error synchronization sequence.  
% \end{enumerate}
% Details are as follows.

{\it Augmented codebook and synchronization sequence.} For any $G$ with $C(G)>0$, there exist two distinct vertices $\b,\g \in \Xc$ such that $(\b,\g) \notin E$. Let $\Cc^{(t)}$ be a zero error channel code of length $t$ and rate $R_\text{ch} = \frac{1}{t}\log{|\Cc^{(t)}|}$. We create an augmented codebook by sandwiching each codeword between two guarding $\g$'s, i.e., 
\[
 \tilde{\Cc}^{(t+2)} = \{(\g,c^t,\g)\suchthat c^t \in \Cc^{(t)}\}.
\]
The sequence 
\[
\b^{t+2} = (\underbrace{\b,\ldots,\b}_{t+2})
\]
will be used as the synchronization sequence.

{\it Phase detection sequence design.} We take a de Bruijn sequence with alphabet size $|\Cc^{(t)}|$ and order $r$. We associate each symbol in the de Bruijn alphabet with a different codeword in the augmented codebook $\tilde{\Cc}^{(t+2)}$ (note that $|\tilde{\Cc}^{(t+2)}| = |\Cc^{(t)}|$ by design). Then, similar to the concatenated structure in Figure~\ref{fig:prob-code}, we concatenate (in a sequential manner) the augmented codewords according to the de Bruijn sequence. Between every $r$ consecutive blocks of augmented codewords, we insert a synchronization block $\b^{t+2}$. This way, the de Bruijn sequence of length $|\Cc^{(t)}|^r$ is mapped to a phase detection sequence $x^n$ of length $n = \frac{(r+1)(t+2)}{r}|\Cc^{(t)}|^r$.

{\it Detection.} We choose the length of the detection window to be
\[
 k = (r+2)(t+2).
\]
This ensures that the window will contain $r$ complete codeword blocks and one complete synchronization block. For each $w_1 \in\{0\} \cup [k-t-2]$, define 
\[
\Sc_y(w_1) \triangleq \left\{u^{t+2} \in \Xc^{t+2}\suchthat \prod_{i=1}^{t+2}p_{Y|X}(y_{w_1+i}|u_i) >0\right\}
\]
as the set of input sequences that may result in the output sequence $y_{w_1+1}^{w_1+t+2}$. The detector finds a $\wh_1 \in \{0\} \cup [k-t-2]$ such that $\b^{t+2} \in \Sc_y(w_1)$. If there are more than one, it chooses the smallest index. It declares an error if there is none. If $\wh_1$ is found, then the first complete block starts from index $(\wh_1 \mod t+2)+1$ of $y^k$. Knowing the block boundary, the detector can then decode the $r$ codewords from $\Cc^{(t)}$. This corresponds to $r$ contiguous symbols in the de Bruijn sequence, which uniquely determine the starting position $\wh_2+1$ in the de Bruijn sequence. Then, the phase estimate is declared as
\[
 \mh = \left(\wh_2 + \left\lceil \frac{\wh_2}{r}\right\rceil\right)(t+2) + 1 - (\wh_1 \mod t+2).
\]

{\it Error Analysis.} The crucial part of the error analysis is to show the synchronization sequence $\b^{t+2}$ can be detected with zero error. Once the block boundary is found, the $r$ codewords can be decoded with zero error, and the location of the corresponding $r$ symbols in the de Bruijn sequence can also be found with zero error.

To see the scheme ensures zero error detection of $\b^{t+2}$, we show that for all $m \in [n]$ such that $x_{m}^{m+t+1} \ne \b^{t+2}$, $(x_{m}^{m+t+1},\b^{t+2}) \notin E_{t+2}$, where $E_{t+2}$ is the edge set of $G^{t+2}$. There are two cases. When $x_{m}^{m+t+1}$ is a complete block, we have $x_m = x_{m+t+1} = \g$ and hence $(x_{m}^{m+t+1}, \b^{t+2}) \notin E_{t+2}$ since $(\b,\g) \notin E$. When $x_{m}^{m+t+1}$ consists of two (partial) blocks, we know at least one block must be a codeword block. Thus, considering the last symbol of the first block and the first symbol of the second block, we know at least one is $\g$. This implies $(x_{m}^{m+t+1},\b^{t+2}) \notin E_{t+2}$. In summary, $\b^{t+2}$ is never confusable with other $x_{m}^{m+t+1}$ at the detector.

{\it Rate.} By design, the rate of the scheme is
\begin{align*}
 R  &= \frac{\log\left(\frac{(r+1)(t+2)}{r}|\Cc^{(t)}|^r\right)}{(r+2)(t+2)}\\
 &= \frac{r\log |\Cc^{(t)}|}{(r+2)(t+2)} + \frac{\log\left(\frac{(r+1)(t+2)}{r}\right)}{(r+2)(t+2)}\\
 &= \frac{r\,t}{(r+2)(t+2)} R_\text{ch} + \frac{\log\left(\frac{(r+1)(t+2)}{r}\right)}{(r+2)(t+2)},
\end{align*}
which tends to $R_\text{ch}$ as $r \to \infty$ and $t \to \infty$. Therefore, by choosing a zero error capacity achieving channel code, the phase detection scheme achieves $C(G)$.
\end{IEEEproof}

\begin{remark}\label{rmk:last-one}
  Suppose $C(G)$ is achieved by a finite block code of length $s>1$ (e.g., for the pentagon graph \cite{Lovasz1979}). In this case, generating a phase detection sequence using a de Bruijn sequence with codewords of the capacity achieving code as symbols, cannot work. To see this, recall that the induced codebook associated with any zero error phase detection sequence forms a zero error channel code of the same rate. However, a simple calculation shows that this rate is equal to $\frac{\log{n}}{\log{(n\slash s)}}\cdot C(G)$, which exceeds the capacity $C(G)$ for $s>1$. 
\end{remark}

\section{Multiple Access: Probabilistic Noise,\\ Vanishing Error}
\label{sec:mac-prob}

So far in all the models we have discussed, phase detection either achieves the best known achievable rate of its channel coding counterpart, or shares the same capacity as that of channel coding. In this section, we encounter the first model, the multiple access phase detection with vanishing error, whose capacity region is strictly included in that of its channel coding counterpart. 

\subsection{Fundamental Limit}
\label{sec:mac-prob-limit}

\begin{theorem}
\label{thm:prob}
 The vanishing error capacity region $\Cr_\text{ve}$ for phase detection over the channel $p(y|x_1,x_2)$ is the set of all rate pairs $(R_1,R_2)$ such that 
 \begin{equation}\label{eqn:mac-ve-limit}
 \begin{split}
  R_1 &\le I(X_1;Y|X_2),\\
  R_2 &\le I(X_2;Y|X_1),\\
  R_1+R_2 &\le I(X_1,X_2;Y)
 \end{split}%
 \end{equation}
 for some $p(x_1)p(x_2)$.
\end{theorem}

\begin{remark}
\label{rmk:gap}
 We note that this region is not convex in general. Compared to the usual MAC capacity region, which is the convex hull of $\Cr_\text{ve}$, 
%  or equivalently, the set of rate pairs $(R_1,R_2)$ such that
%  \begin{align*}
%   R_1 &\le I(X_1;Y|X_2,Q),\\
%   R_2 &\le I(X_2;Y|X_1,Q),\\
%   R_1+R_2 &\le I(X_1,X_2;Y|Q)
%  \end{align*}
%  for some $p(q)p(x_1|q)p(x_2|q)$ with cardinality bound on $Q$ as $|\Qc| \le 2$. 
 this region can be a strict subset (see, for example, the \emph{push-to-talk} MAC with binary inputs and output, given by $p(0|0,0) = p(1|0,1) = p(1|1,0) = 1$ and $p(0|1,1) = 1/2$~\cite[Problem 3.2.6]{Csiszar--Korner1981}). 
\end{remark}

\begin{IEEEproof}[Proof of Theorem~\ref{thm:prob}] We prove the achievability through random sequence generation and joint typicality detection.

\smallskip
{\it Sequence generation.} Fix a pmf $p(x_1)p(x_2)$. Let $n_1 = 2^{kR_1}$ and $n_2 = 2^{kR_2}$. We generate the two sequences $X_1^{n_1}$ i.i.d.$~\sim p(x_1)$ and $X_2^{n_2}$ i.i.d.$~\sim p(x_2)$. 

\smallskip
{\it Detection.} Upon receiving $y^{k}$, the detector declares the phase estimate $(\mh_1, \mh_2) \in [n_1]\times [n_2]$ if it is the unique pair such that $(\phi_1(\mh_1),\phi_2(\mh_2),y^k) \in \Tc_\e^{(k)}(X_1,X_2,Y)$; if there is none or more than one, it declares an error. 

\smallskip
{\it Analysis of the probability of error.}
Without loss of generality, we assume that the correct phase pair is $(M_1,M_2) = (1,M_2)$. the detector makes an error only if one or more of the following events occur:
\begin{align*}
 \Ec_1 &= \{(\phi_1(1),\phi_2(M_2),Y^k) \notin \Tc_\e^{(k)}\},\\
 \Ec_2 &= \{(\phi_1(m_1),\phi_2(M_2),Y^k) \in \Tc_\e^{(k)} \text{ for some } m_1 \neq 1\},\\
 \Ec_3 &= \{(\phi_1(1),\phi_2(m_2),Y^k) \in \Tc_\e^{(k)} \text{ for some } m_2 \neq M_2\},\\
 \Ec_4 &= \{(\phi_1(m_1),\phi_2(m_2),Y^k) \in \Tc_\e^{(k)} \\
 &\hspace{8em}\text{ for some } m_1 \neq 1 \text{ and } m_2 \neq M_2\}.
\end{align*}
By the law of large number, $\P(\Ec_1)$ tends to zero as $k \to \infty$. For $\Ec_2$, we have
\begin{align*}
 &\P(\Ec_2) \le \sum_{m_1 = k+1}^{n_1-k+1} \P\{(\phi_1(m_1),\phi_2(M_2),Y^k) \in \Tc_\e^{(k)}\}\\
 & + \sum_{m_1 = 2}^{k} \P\{(\phi_1(m_1),\phi_2(M_2),Y^k) \in \Tc_\e^{(k)}\}\\
 & + \sum_{m_1 = n_1-k+2}^{n_1} \P\{(\phi_1(m_1),\phi_2(M_2),Y^k) \in \Tc_\e^{(k)}\}\\
 &\stackrel{(a)}{\le} (2^{kR_1}-2k+1)2^{-kI(X_1;Y|X_2)} +  2(k-1)2^{-k\gamma_1(\e)},
\end{align*}
which tends to zero as $k \to \infty$ if $R_1 < I(X_1;Y|X_2) -\d(\e)$.
Here the first term in ($a$) follows since $\phi_1(m_1)$ in that range does not overlap with the right chunk $\phi_1(1)$ and the probability can be bounded by the packing lemma~\cite[Lemma~3.1]{El-Gamal--Kim2011}. The second term in ($a$) corresponds to the overlapping chunks and the probability is bounded by Lemma~\ref{lem:iid}  with $X_{m}^{m+k-1} \leftarrow \phi_1(m_1), Y^k \leftarrow (\phi_2(M_2),Y^k)$.
We can similarly show that $\P(\Ec_3)$ tends to zero as $k \to \infty$ if $R_2 < I(X_2;Y|X_1) - \d(\e)$. For $\Ec_4$, there are four different cases:
\begin{itemize}
 \item There are $(2^{kR_1}-2k+1)(2^{kR_2}-2k+1)$ pairs $(m_1,m_2)$ such that neither $\phi_1(m_1)$ nor $\phi_2(m_2)$ overlaps with the right chunks. By the packing lemma, we can bound
\begin{align*}
 &\P\{(\phi_1(m_1),\phi_2(m_2),Y^k) \in \Tc_\e^{(k)} \}\le 2^{-kI(X_1,X_2;Y)}.
\end{align*}
  \item There are $2(k-1)(2^{kR_2}-2k+1)$ pairs $(m_1,m_2)$ such that $\phi_1(m_1)$ overlaps with the right chunk while $\phi_2(m_2)$ does not. Applying Lemma~\ref{lem:iid} first and then the packing lemma (note the independence between $\phi_1(m_1)$ and $\phi_2(m_2)$ for any $m_1$ and $m_2$), we have 
%  \begin{align*}
%   &\P\{(\phi_1(m_1),\phi_2(m_2),Y^k) \in \Tc_\e^{(k)} \}\\
%   &=\P\{(\phi_1(m_1),Y^k )\in \Tc_\e^{(k)}\}\\
%   &\hspace{1em}\cdot\P\{(\phi_1(m_1),\phi_2(m_2), Y^k) \in \Tc_\e^{(k)}\\
%   &\hspace{10em}\big|(\phi_1(m_1),Y^k) \in \Tc_\e^{(k)}\}\\
%   &\le 2^{-k(\gamma_2(\e)+I(X_2;Y|X_1))}.
%  \end{align*}
 \begin{align*}
  &\P\{(\phi_1(m_1),\phi_2(m_2),Y^k) \in \Tc_\e^{(k)} \}\\
  &=\P\{(\phi_1(m_1),Y^k )\in \Tc_\e^{(k)}\}\\
  &\hspace{.6em}\cdot\P\{(\phi_1(m_1),\phi_2(m_2), Y^k) \in \Tc_\e^{(k)}\big| \\
  &\hspace{13em}(\phi_1(m_1),Y^k)\in \Tc_\e^{(k)}\}\\
  &\le 2^{-k(\gamma_2(\e)+I(X_2;Y|X_1))}.
 \end{align*}
 \item There are $2(k-1)(2^{kR_1}-2k+1)$ pairs $(m_1,m_2)$ such that $\phi_2(m_2)$ overlaps with the right chunk while $\phi_1(m_1)$ does not. Similarly, we have
 \begin{align*}
  &\P\{(\phi_1(m_1),\phi_2(m_2),Y^k) \in \Tc_\e^{(k)} \}\\
  &\le 2^{-k(\gamma_3(\e)+I(X_1;Y|X_2))}.
 \end{align*}
 \item The rest $4(k-1)^2$ pairs are such that both $\phi_1(m_1)$ and $\phi_2(m_2)$ overlap with the right chunks. We note that the event $\{(\phi_1(m_1),\phi_2(m_2),Y^k) \in \Tc_\e^{(k)}(X_1,X_2,Y)\}$ implies $\{(\phi_1(m_1),Y^k) \in \Tc_\e^{(k)}(X_1;Y)\}$. Thus, we can bound the error as
 \begin{align*}
  &\P\{(\phi_1(m_1),\phi_2(m_2),Y^k) \in \Tc_\e^{(k)} \}\\
  &\le \P\{(\phi_1(m_1),Y^k) \in \Tc_\e^{(k)} \}\\
  &\le 2^{-k\gamma_2(\e)}.
 \end{align*}
\end{itemize}
Combining all four cases, we have $\P(\Ec_4)$ tends to zeros as $k \to \infty$ if $R_1+R_2 < I(X_1,X_2;Y) - \d(\e)$, $R_1 < I(X_1;Y|X_2) -\d(\e)$, and $R_2 < I(X_2;Y|X_1)-\d(\e)$. Letting $\e \to 0$ completes the proof of the achievability.

\smallskip
For the converse, we wish to show for any $(2^{kR_1},2^{kR_2},k)$ multiple access phase detection scheme with vanishing probability of error $\lim_{k \to \infty}P_e^{(k)} = 0$, the rate pair $(R_1,R_2) \in \Cr_\text{ve}$. Given the two sequences $x_1^{n_1}$ and $x_2^{n_2}$, the joint distribution of $(M_1,M_2,Y^k)$ is
 \[
  \frac{1}{2^{k(R_1+R_2)}}\prod_{i=0}^{k-1} p_{Y|X_1,X_2}(y_{1+i}|x_{1,m_1+i},x_{2,m_2+i}). %2^{-k(R_1+R_2)}
 \]
 By Fano's inequality, we have $H(M_1,M_2|Y^k)\le k(R_1+R_2)P_e^{(k)} + 1 \le k\e_k$,
 where $\e_k$ tends to zero as $k \to \infty$. We bound the sum rate as follows
 \begin{align*}
  &k(R_1+R_2)\\
  &= H(M_1,M_2)\\
  &\stackrel{(a)}{\le} I(M_1,M_2;Y^k) + k\e_k\\
  &= \sum_{i=0}^{k-1} I(M_1,M_2;Y_{1+i}|Y^i) + k\e_k\\
  &\stackrel{(b)}{\le} \sum_{i=0}^{k-1} I(M_1, M_2,Y^i, x_{1,M_1+i},x_{2,M_2+i};Y_{1+i}) + k\e_k\\
  &\stackrel{(c)}{=} \sum_{i=0}^{k-1} I(x_{1,M_1+i},x_{2,M_2+i};Y_{1+i}) + k\e_k,
 \end{align*}
where ($a$) follows from Fano's inequality and ($c$) follows since $(M_1,M_2,Y^{i}) \to (x_{1,M_1+i},x_{2,M_2+i}) \to Y_{1+i}$ form a Markov chain due to the memorylessness of the channel. Note here in both ($b$) and ($c$), $x_{j,M_j+i}$ is a function of $M_j$ that takes value $x_{j,m_j+i}$ when $M_j = m_j$ for $j = 1,2$. Now we bound the individual rate as follows
\begin{align*}
 kR_1 &= H(M_1|M_2)\\
 &\stackrel{(c)}{\le} I(M_1;Y^k|M_2) + k\e_k\\
 &= \sum_{i=0}^{k-1} I(M_1;Y_{1+i}|M_2,Y^i) + k\e_k\\
 &\le \sum_{i=0}^{k-1} I(Y^i,M_1,M_2,x_{1,M_1+i};Y_{1+i}|x_{2,M_2+i}) + k\e_k\\
 &= \sum_{i=0}^{k-1} I(x_{1,M_1+i};Y_{1+i}|x_{2,M_2+i}) +k\e_k,
\end{align*}
where ($c$) follows since $H(M_1|Y^k,M_2) \le H(M_1,M_2|Y^k) \le k\e_k$. Now flipping the role of $1$ and $2$, we have
\[
kR_2 \le \sum_{i=0}^{k-1} I(x_{2,M_2+i};Y_{1+i}|x_{1,M_1+i}) +k\e_k.
\]
Now we introduce a time-sharing random variable $Q \sim \U[k]$, which is independent of $(M_1,M_2,Y^k)$. We can write
\begin{align*}
 R_1+R_2 &\le  I(x_{1,M_1+Q},x_{2,M_2+Q};Y_{1+Q}|Q) + \e_k,\\
 R_1 &\le  I(x_{1,M_1+Q};Y_{1+Q}|x_{2,M_2+Q},Q) + \e_k,\\
 R_2 &\le  I(x_{2,M_2+Q};Y_{1+Q}|x_{1,M_1+Q},Q) + \e_k.
\end{align*}
Note that $\P\{Y_{1+Q}=y|x_{1,M_1+Q} = x_1,x_{2,M_2+Q} = x_2\} = p(y|x_1,x_2)$, which is distributed according to the channel conditional pmf. Hence, we identify $X_1 = x_{1,M_1+Q}, X_2 = x_{2,M_2+Q}$, and $Y = Y_{1+Q}$ to  obtain
\begin{align*}
 R_1+R_2 &\le  I(X_1,X_2;Y|Q) + \e_k\\
 &\le I(Q,X_1,X_2;Y) + \e_k\\
 &\stackrel{(d)}{=} I(X_1,X_2;Y) + \e_k,
\end{align*}
where $(d)$ follows since $Q \to (X_1,X_2) \to Y$ form a Markov chain. We similarly obtain
\begin{align*}
 R_1 &\le I(X_1;Y|X_2) + \e_k,\\
 R_2 &\le I(X_2;Y|X_1) + \e_k.
\end{align*}
Note that since $M_1$ and $M_2$ are independent and uniform over $[n_1]$ and $[n_2]$, $M_1 + Q$ and $M_2+Q$ are independent, and so are $x_{1,M_1+Q}$ and $x_{2,M_2+Q}$. Therefore, we can restrict the inputs to independent distribution $p(x_1)p(x_2)$.
Letting $k \to \infty$ completes the proof of the converse. 
\end{IEEEproof}

\begin{remark}
 We note the connection between the above converse proof, and that of the totally asynchronous MAC~\cite{Poltyrev1983, Hui--Humblet1985}. Unlike channel coding in the usual (synchronous) MAC setting, where the two inputs can be correlated through the time-sharing random variable $Q$, the two inputs $x_{1,M_1+Q}$ and $x_{2,M_2+Q}$ in the phase detection setting are independent even with the time-sharing random variable. Therefore, while the input pmf for the channel coding problem is $p(q)p(x_1|q)p(x_2|q)$, it is $p(x_1)p(x_2)$ in the phase detection setting. This essentially results in the strict gap between the capacity regions of the two problems (see also Remark~\ref{rmk:gap}).
\end{remark}

\begin{remark}
 One can similarly show that the vanishing error capacity region for phase detection in the $L$-user MAC $p(y|x_1,\ldots,x_L)$ is the set of rate tuples $(R_1,\ldots,R_L)$ such that
 \begin{equation}
 \label{eqn:lmac-prob}
  \sum_{i \in \Jc} R_i \le I(X_\Jc;Y|X_{\Jc^c}) \quad \text{for every } \Jc \subseteq [L]
 \end{equation}
for some $\prod_{i=1}^L p(x_i)$. Here $X_\Jc = \{X_i\suchthat i\in\Jc\}$.
\end{remark}

% 
% \lele{Shall we somehow remark on totally asynchronous MAC, or is it a detour? If we do, we need to define it first, and somehow explain it is different from our problem---a code for the phase detection does not directly translate to a totally asynchronous MAC code. But somehow they have the same capacity region.}

\subsection{A Low-Complexity Construction}
\label{sec:mac-prob-code}

In this section, we build on the point-to-point phase detection sequence construction in Section~\ref{sec:prob-code} and provide an $O(k\log{k})$ complexity sequence construction that achieves any rate pairs $(R_1,R_2) \in \Cr_\text{ve}$. The construction consists of several ingredients:
\begin{enumerate}
\item the vanishing error capacity achieving phase detection sequence for a point-to-point channel $p(y|x)$, as given in Section~\ref{sec:prob-code},

\item the rate-splitting~\cite{Grant--Rimoldi--Urbanke--Whiting2001} technique, which is a point-to-point channel coding technique for achieving arbitrary rate pairs in the  MAC region without time sharing, and

\item a novel symbol-by-symbol mapping that enables rate-splitting in the phase detection setting.
\end{enumerate}
Details are as follows.

\smallskip

{\it Rate splitting.} In the random coding scheme, we simultaneously detect the phases $m_1$ and $m_2$ by checking typicality of all possible pairs via brute force. In practice, it is unclear whether simultaneous detection can be implemented at low complexity. In our design, we circumvent this difficulty by employing the rate splitting technique of~\cite{Grant--Rimoldi--Urbanke--Whiting2001}, which transforms the MAC coding problem into three point-to-point channel coding problems. Fix a pmf $p(u)p(v)p(x_2)$ and a function $x_1(u,v)$. We target the rate pair 
\begin{align}
\label{eqn:split-rate}
\begin{split}
R_1 &= I(U;Y) + I(V;Y|X_2,U),\\
R_2 &= I(X_2;Y|U).
\end{split}
\end{align}
It is known~\cite{Grant--Rimoldi--Urbanke--Whiting2001} that for any rate point $(I_1,I_2) \in \Cr_\text{ve}$, there exists a pmf $p(u)p(v)p(x_2)$ and a function $x_1(u,v)$ such that $I_1 = R_1$ and $I_2 = R_2$.

% Note that for any choice of $(U,V)$, $R_1 + R_2 = I(U,V,X_2;Y) = I(X_1,X_2;Y)$. Choosing $U = \emptyset, V = X_1$, we have $R_1 = I(X_1;Y|X_2), R_2 = I(X_2;Y)$, which corresponds to a corner point in $\Cr_\text{ve}$. Considering the other extreme $U = X_1,V = \emptyset$, we have $R_1 = I(X_1;Y), R_2 = I(X_2;Y|X_1)$, which corresponds to the other corner point. Therefore, by the continuity of mutual informations, for any rate point in $\Cr_\text{ve}$, there exists a \emph{split} of $X_1$ into $(U,V)$ such that the target rate pair achieves it. 

\smallskip

{\it Sequence construction.} We design three vanishing error phase detection sequences for three point-to-point channels $U \to Y$, $X_2 \to (Y,U)$, and $V \to (Y,U,X_2)$ respectively, each according to the construction in Section~\ref{sec:prob-code}. Specifically, $u^{n_u}$, $x_2^{n_2}$, and $v^{n_v}$ are $(I(U;Y),l,t,t_u)$, $(I(X_2;Y|U),l,t,t_2)$, and $(I(V;Y|U,X_2),l,t,t_v)$ point-to-point phase detection sequences, respectively (see Remark~\ref{rmk:p2p-code} for the definition of an $(R_\text{ch},l,t,\tau)$ point-to-point phase detection sequence). 

Given $u^{n_u}$ and $v^{n_v}$, we form an $x_1$ sequence of length $n_u n_v$ through the symbol-by-symbol mapping 
\begin{equation}
\label{eqn:mapping}
x_{1,m_1} = x_1(u_{m_u},v_{m_v}) \text{ for } m_1 \in [n_un_v], 
\end{equation}
where 
\begin{align*}
m_u &= m_1 \pmod{n_u},\\
m_v &= m_1\pmod{n_v}. 
\end{align*}
Note that when $n_u$ and $n_v$ are relatively prime, each phase $m_1 \in [n_un_v]$ corresponds to a \emph{distinct} phase pair $(m_u,m_v)$. Moreover, the way the $u$ and the $v$ sequences are ordered ensures that any length-$k$ chunk of the $x_1$ sequence is formed from a length-$k$ chunk of the $u$ sequence and a length-$k$ chunk of the $v$ sequence. Such an $x_1^{n_un_v}$ sequence simulates the channel output when the two phase detection sequences $u^{n_u}$ and $v^{n_v}$ go through a deterministic MAC $x_1(u,v)$. Finally, recall that any $t_u$ (and $t_v$)  that is a linear function of $t$ results in the same asymptotic rate of the phase detection sequence (cf. Section~\ref{sec:prob-code}). Hence, by adjusting the parameters $t_u$ and $t_v$, it is always possible to make $n_u$ and $n_v$ relatively prime.

\smallskip
{\it Detection.} The way the $u,x_2,v$ sequences  are designed allows multiple access phase detection through successive point-to-point phase detection in the channels $U \to Y$, $X_2 \to (Y,U)$, and $V \to (Y,U,X_2)$. We choose the length of the detection window to be
\[
k = lt + 3\max\{t_u,t_v,t_2\} + \max\{t,t_u,t_v,t_2\}
\]
and successively detect the phases in the order $\mh_u \to \mh_{2} \to \mh_v$. The phase of the $x_1$ sequence is declared to be the unique $\mh_1 \in [n_un_v]$ such that $\mh_1 \pmod{n_u} = \mh_u$ and $\mh_1 \pmod{n_v} = \mh_v$.

\smallskip
{\it Analysis of the probability of error.} By the analysis in the point-to-point case, the probability of error for detecting each sequence $\P(\Ec_j)$, $j = 1,2,3$, tends to zero as $t \to \infty$. By successive cancellation, the total probability of error $\P(\Ec) \le \P(\Ec_1) + \P(\Ec_2) + \P(\Ec_3)$, which tends to zero as $t \to \infty$.

\smallskip
{\it Rate.} Letting $t \to \infty$ and then $l \to \infty$, the rates of the $u$, $x_2$, and $v$ phase detection sequences approach, respectively,  
\begin{align*}
 R_u &= I(U;Y),\\
 R_2 &= I(X_2;Y|U),\\
 R_v &= I(V;Y|U,X_2).
\end{align*}
Moreover, we have
\begin{align*}
R_1 &= \frac{\log{n_1}}{k} \\
&= \frac{\log{n_un_v}}{k}\\
&= \frac{\log{n_u}+\log{n_v}}{k}\\
&= R_u + R_v \\
&= I(U;Y) + I(V;Y|U,X_2),
\end{align*}
which, together with $R_2$, is exactly our target rate pair~\eqref{eqn:split-rate}.

\smallskip
{\it Complexity.} Each of the three point-to-point phase detection sequence has detection complexity $O(k\log{k})$. Therefore, the total complexity of the multiple access detection complexity is also $O(k\log{k})$.

\begin{remark}
 The original symbol-by-symbol mapping in the channel coding setting, which maps
 \[
  x_{1i} = x_1(u_i,v_i),
 \]
does not provide the desired relation $R_1 = R_u + R_v$ in the sequence setting. This is because knowing the phase $m_u$ simultaneously reveals the phase $m_v$. In contrast, the mapping in~\eqref{eqn:mapping} ensures that for each phase $m_u \in [n_u]$, all possible phases $m_v \in [n_v]$ appear in the $x_1$ sequence. This creates the independence between the two phases $M_u$ and $M_v$.
\end{remark}

\begin{remark}
 Rate splitting can be generalized to $L$-user MACs. More precisely, one can split $X_j$, $j \in[L-1]$, into two auxiliary layers $U_j$ and $V_j$, and keep $X_L$ unsplit. It is shown~\cite{Grant--Rimoldi--Urbanke--Whiting2001} that there exists a successive decoding order that achieves any rate tuple in the $L$-user MAC region~\eqref{eqn:lmac-prob}. Together with the symbol-by-symbol mappings $x_j(u_j,v_j)$ applied as in~\eqref{eqn:mapping}, we can design a rate-optimal low-complexity phase detection scheme for an $L$-user MAC.
\end{remark}

\section{Multiple Access: Probabilistic Noise,\\ Zero Error}
\label{sec:mac-advs}

% So far, we have seen that phase detection achieves the same best known rate as channel coding in the adversarial point-to-point setting (Section~\ref{sec:advs}), and the same capacity in the probabilistic point-to-point setting (Section~\ref{sec:prob}). For the later setup, a code-level equivalence between the two settings is established through the construction in Section~\ref{sec:prob-code} (see Remarks~\ref{rmk:eqvt-avg} and~\ref{rmk:eqvt-max}). It might be tempting to expect a similar equivalence between the two settings in multiple access channels. However, as we will see in Sections~\ref{sec:separation} and~\ref{sec:mac-prob-limit}, the capacity region of phase detection can be strictly included in that of channel coding, with 
In this section, we consider zero error phase detection in multiple access channels. We first demonstrate a strict separation between the channel coding setting and the phase detection setting in Section~\ref{sec:separation}. Then, we restrict our attention to zero error phase detection in the modulo-2 addition MAC in Sections~\ref{sec:mac-advs-limit} and~\ref{sec:mac-advs-code}. We note that for channel coding in the modulo-2 addition MAC, any rate pair in the zero-error capacity region $\{(R_1,R_2)\suchthat R_1 + R_2 \le 1\}$ can be achieved by time sharing between two rate-one codes. However, time sharing is not applicable in the phase detection scenario. Thus, our sequence design requires different ideas.
% where the phase pair $(m_1,m_2)$ needs to be recovered with zero error. 

\subsection{Separation Between Phase Detection and Channel Coding}
\label{sec:separation}

Let us consider again the push-to-talk MAC (see definition in Remark~\ref{rmk:gap}). The zero error capacity region for channel coding is the set of rate pairs $(R_1,R_2)$ such that 
\begin{equation}
\label{eqn:push-to-talk}
 R_1 + R_2 \le 1.
\end{equation}
To see this, first note that the two corner points $(0,1)$ and $(1,0)$ can be achieved with zero error using any channel code of rate 1, and other points are achievable by time sharing. Moreover, since the output alphabet is binary, the rate region~\eqref{eqn:push-to-talk} is also an outer bound.

For zero error phase detection, a simple outer bound of $\Cr_\text{ze}$ is its vanishing error counterpart $\Cr_\text{ve}$, which is shown to be the rate region~\eqref{eqn:mac-ve-limit} in Theorem~\ref{thm:prob}. For any rate pair $(R_1,R_2)$ in the rate region~\eqref{eqn:mac-ve-limit},
\begin{align*}
 R_1 + R_2 &\le I(X_1,X_2;Y)\\
 & = H(Y) - H(Y|X_1,X_2)\\
 &\stackrel{(a)}{\le} 1 - p_{X_1}(1)p_{X_2}(1)\\
 &\stackrel{(b)}{\le} 1.
\end{align*}
For equalities in both ($a$) and ($b$) to hold, we must have $p_{X_1}(1) = 0, p_{X_2}(1) = 1/2$ or $p_{X_1}(1) = 1/2, p_{X_2}(1) = 0$, which correspond to the two corner points $(0,1)$ and $(1,0)$ respectively. Any other input pmf $p(x_1)p(x_2)$ results in a sum rate strictly less than 1. Therefore, other than the two corner points, the rate pair $(R_1,R_2)$ along the line $R_1 + R_2 =1$ is not achievable in the phase detection setting, which establishes the separation.

\subsection{Fundamental Limit for Modulo-2 Addition MAC}
\label{sec:mac-advs-limit}
\begin{theorem}
 \label{thm:mac-advs}
The zero error capacity region $\Cr_\text{ze}$ for multiple access phase detection over the channel $Y = X_1 \oplus X_2$ is the set of rate pairs $(R_1,R_2)$ such that %$R_1 + R_2 \le 1$.
 \[
  R_1 + R_2 \le 1.
 \]
\end{theorem}

\begin{IEEEproof}
As argued above, this rate region is an outer bound since any codebooks induced by the zero error phase detection scheme can be used for the zero error channel coding problem. In what follows, we prove the achievability of this region using properties of linear codes, in a way that resembles Wyner's linear code for the Slapian--Wolf problem~\cite{Wyner1974}.

We choose the first sequence $x_1^{n_1}$ to be a linear sequence generated by LFSR with a primitive characteristic polynomial $a(z) = \sum_{i=0}^{r-1} a_i z^i + z^r$ over $GF(2)$ (cf. Section~\ref{sec:advs-code}). Then, for $r \le k \le n_1$, the induced codebook together with the all-zero codeword $\Cc_1 \cup \{0^k\}$ form a linear code. Let $H_{(k-r)\times k}$ be a parity check matrix of this linear code. This allows us to define $2^{k-r}$ \emph{cosets}
\[
 \Cc(s^{k-r}) = \{a^k\suchthat H a^k = s^{k-r}\} \subseteq \{0,1\}^k.
\]
Clearly, the linear code belongs to the zero coset $\Cc_1\cup \{0^k\} = \Cc(0^{k-r})$. 

Now, suppose there exist a sequence $x_2^{n_2}$ such that each length-$k$ chunk $\phi_2(m_2)$ of the sequence belongs to a different non-zero coset $\Cc(s^{k-r})$ with $s^{k-r} \neq 0^{k-r}$. Then, the phase pair $(m_1,m_2)$ can be recovered with zero-error from their sum through successive cancellation detection as follows. We take $H(\phi_1(m_1) \oplus \phi_2(m_2)) = 0^{k-r} \oplus H\phi_2(m_2) \triangleq s^{k-r}$. By design, there is only one chunk of $x_2^{n_2}$ that belongs to the coset $\Cc(s^{k-r})$. This uniquely determines the phase $m_2$. Once $\phi_2(m_2)$ is recovered, we know $\phi_1(m_1) = y^k \oplus \phi_2(m_2)$. Then by design, $m_1$ can be uniquely determined by its first $r$ bits. 

We now proceed to show the existence of such a sequence $x_2^{n_2}$, using Lov\'{a}sz local lemma (Lemma~\ref{lem:lll}). We generate $X_2^{n_2}$ i.i.d. uniform. Let the ``bad'' events be $A_j = \{H \phi_2(m_2) = H\phi_2(m_2') \text{ for some } m_2 \neq m_2'\}$. Since $\phi_2(m_2)\oplus \phi_2(m_2')$ is i.i.d. uniform whether or not the two chunks overlap, the probability that the sum falls in the null space of $H$ is
\[
 \P(A_j) = \P\{H(\phi_2(m_2)\oplus \phi_2(m_2')) = 0^{k-r}\} = \frac{1}{2^{k-r}}.
\]
Now each $A_j$ is mutually independent of all other events, except for a set of at most $4kn_2$ events. This is because the random variable $\phi_2(m_2)\oplus \phi_2(m_2')$ is mutually independent of all $X_{2i}$'s with $i \in [n_2]\setminus \{m_2-k+1,m_2-k+2,\ldots,m_2+k-1\} \setminus \{m_2'-k+1,m_2'-k+2,\ldots,m_2'+k-1\}$, which excludes at most $4kn_2$ events. Applying Lemma~\ref{lem:lll}, the sequence $X_2^{n_2}$ exists with positive probability
\[
 \P\left\{\cap_{j=1}^{\frac12 n_2(n_2-1)}A_j^c\right\}>0
\]
if
\[
 16kn_22^{-(k-r)} \le 1,
\]
or equivalently
\[
 \frac{\log{n_2}}{k} + \frac{r}{k} \le 1-\frac{\log{(16k)}}{k}.
\]
By the definition of the rates, $R_2 = \frac{\log{n_2}}{k}$ and $R_1 = \frac{\log{n_1}}{k} = \frac{\log(2^r-1)}{k}\approx \frac{r}{k}$.
Letting $k\to \infty$, we conclude that a good sequence $x_2^{n_2}$ exists if $R_1 + R_2 < 1$. 
\end{IEEEproof}

% \begin{remark}
%  It might be temping to randomize both sequences $X_1^{n_1}$ and $X_2^{n_2}$, but only achieves $(1/3,1/3)$...
% \end{remark}
% 
% \begin{remark}
%  Another way to design the sequences is to let $x_1^{n_1}$ be a de Bruijn sequence such that every length-$k$ chunk has weight less than $pk$ (see~\cite{}), and let $x_2^{n_2}$ be a sequence such that the set of length-$k$ chunks has minimum distance $d > 2pk$. The corresponding achievable rate region is 
% $ \left(\cup_{0\le p\le 1/4} \{(R_1,R_2)\suchthat R_1 \le p, R_2 \le 1-h(2p)\}\right)\bigcup$ $ \left(\cup_{0\le p\le 1/4} \{(R_1,R_2)\suchthat R_1 \le 1-h(2p), R_2 \le p\}\right)$, which is a strict subset of the capacity region above.
% \end{remark}
% 

\subsection{Sequence Construction for Modulo-2 Addition MAC}
\label{sec:mac-advs-code}

In this Section, we show that not only does the sequence $x_2^{n_2}$ from the previous section exists, but it can also be a linear sequence. Moreover, we provide an explicit sequence construction that achieves any rate pair $(R_1,R_2) \in \Cr_\text{ze}$. 

{\it Sequence Construction.}
Let $a_1(z),a_2(z) \in \Ff_2(z)$ be two distinct primitive polynomials of degree $r_1$ and $r_2$ respectively. Let $x_1^{n_1}$ and $x_2^{n_2}$ be the two linear sequences generated by the $a_1(z)$ and $a_2(z)$ respectively (cf. Section~\ref{sec:advs-code}). Letting $k \ge r_1 + r_2$, the rates of the two sequences are 
\[
 R_1 = \frac{\log(2^{r_1}-1)}{k} \approx \frac{r_1}{k} 
\]
and
\[
R_2 = \frac{\log(2^{r_2}-1)}{k} \approx \frac{r_2}{k}.
\]

\smallskip
{\it Analysis of Detectability.} We need to show that every element in $\Cc_\text{sum}$ can be uniquely expressed as an element in $\Cc_1$ plus an element in $\Cc_2$ (see definitions of $\Cc_1,\Cc_2$, and $\Cc_\text{sum}$ in~\eqref{eqn:c1}, \eqref{eqn:c2} and~\eqref{eqn:csum} respectively). Let us first recall some definitions and facts from the LFSR theory~\cite{Lidl--Niederreiter2008}.

% We say a phase detection sequence $x^n$ is \emph{linear} if the induced codebook together with the zero codeword $\mathcal{C}\cup \{0^k\}$ forms a linear code. %Let $r$ be the dimension of this linear code. Then, the length of the linear phase detection sequence is $n = 2^{r}-1$. 

A one-sided infinite binary sequence $\xv = \{x_i\}_{i \in \mathbb{N}}$ is said to be an \emph{LFSR sequence} if it satisfies the recursion
$ x_{r+j} = \sum_{i=0}^{r-1} a_i x_{i+j}$, for all $j \in \mathbb{N}$.
The polynomial $a(z) = z^r + \sum_{i=0}^{r-1}a_iz^i$ is a \emph{characteristic polynomial} of $\bf x$. The first $r$ bits $(x_1,\ldots,x_r)$ is the \emph{initial state} of $\xv$.
% \footnote{Note the distinction we make between the phase detection sequence $x^n$ and the LFSR sequence $\xv$. The former is a circulant sequence with $x_j = x_{(j-1 \mod n)+1}$. When a phase detection sequence is generated by an LFSR, we do not emphasize which initial state it is generated from as long as the state is nonzero.} 

Let $\Sc(a(z)) = \{\xv\suchthat a(z)$ is a characteristic polynomial of $\xv\}$. Then $\Sc(a(z))$ contains $2^r$ sequences, each corresponding to an $r$-bit initial state. One can check that $\Sc(a(z))$ is an $r$ dimensional vector space over $\Ff_2$. Define $\Sc(a_1(z)) + \Sc(a_2(z))$ $ = \{\xv \oplus \yv\suchthat \xv \in \Sc(a_1(z)), \yv \in \Sc(a_2(z))\}$. When $a_1(z)$ and $a_2(z)$ are relatively prime, we have~\cite[Theorems 8.54, 8.55]{Lidl--Niederreiter2008}
\begin{align}
 \Sc(a_1(z)) \cap \Sc(a_2(z)) &= \{\bf 0\},\label{eqn:intersectoin}\\
 \Sc(a_1(z)) + \Sc(a_2(z))& = \Sc(a_1(z)a_2(z)).\label{eqn:sum}
\end{align}
This is exactly the case for our construction, since $a_1(z)$ and $a_2(z)$ are distinct primitive polynomials and hence relatively prime.

Now suppose that there exists a nonzero $c^k \in \Cc_1 \cap \Cc_2$. One can find an $\xv \in \Sc(a_1(z))$ such that $x^k = c^k$, since the first $r_1$ bits of $\xv$ can be arbitrary and the rest $k-r_1$ bits are generated by the same polynomial $a_1(z)$. Similarly there is a $\yv \in \Sc(a_2(z))$ such that $y^k = c^k$. Note that $\xv \oplus \yv \in \Sc(a_1(z)a_2(z))$ by~\eqref{eqn:sum} and that $a_1(z)a_2(z)$ is a polynomial of degree $r_1+r_2 \le k$. The first $k$ bits $x^k \oplus y^k = 0^k$
fully determines the whole sequence, hence $\xv \oplus \yv = {\bf 0}$. It follows that $\xv = \yv$ and $\xv,\yv \in \Sc(a_1(z)) \cap \Sc(a_2(z))$, which contradicts~\eqref{eqn:intersectoin} since $\xv$ and $\yv$ are nonzero sequences starting with $c^k \neq 0^k$. This proves that $\Cc_1 \cap \Cc_2 = \emptyset$. Notice that $\Cc_i \cup \{0^k\}$ is a linear code, and thus a vector space over $\Ff_2$ (cf. Theorem~\ref{thm:lfsr}). For two vector
spaces $A$ and $B$, we know $\dim(A)+\dim(B)=\dim(A\cap B)+\dim(A+B)$.
Therefore every element in $\Cc_\text{sum}$ can be uniquely
expressed as $c_1^k \oplus c_2^k$, where $c_i^k \in \Cc_i, i= 1,2$.

\begin{remark}
 Here the crucial property is that $a_1(z)$ and $a_2(z)$ are relatively prime. In order to generalize to more than two users, one can choose $L$ distinct primitive polynomials $a_1(z),\ldots,a_L(z)$. This ensures they are relative prime. Thus~\eqref{eqn:intersectoin}~and~\eqref{eqn:sum} generalize as
{\allowdisplaybreaks
\begin{align*}
 \Sc(a_1(z)) \cap \cdots \cap \Sc(a_L(z)) &= \{\bf 0\},\\
 \Sc(a_1(z)) + \cdots +  \Sc(a_L(z))& = \Sc(a_1(z)\cdots a_L(z)).
\end{align*}}%
Constructing the phase detection sequences from these polynomials and following a similar analysis, one can show that the zero error capacity region for phase detection in the $L$-user modulo-2 addition MAC is the set of rate tuples $(R_1,\ldots,R_L)$ such that $\sum_{i=1}^L R_i \le 1$.
\end{remark}

\section{Future Research}
There are several remaining questions. In adversarial point-to-point channel coding, it is easy to show that the GV bound can be attained using linear codes. Is it also true that a linear phase detection scheme can achieve that bound? Furthermore, since sequence design is more difficult than codebook design, can we obtain upper bounds on $C_\text{ad}(p)$ that are tighter than the ones obtained for the adversarial channel coding setup? 

In the zero error point-to-point setup, we have shown how to achieve $C(G)$ in the limit of long sequences. However, when $C(G)=\alpha(G)$, this can be achieved in finite length. Suppose that $C(G)>\alpha(G)$ and is achieved by a finite length channel block code. Does there exists a finite length phase detection sequence of rate exactly $C(G)$? 

% Perhaps the most interesting MAC for practical positioning systems is the binary adder channel $Y = X_1 + X_2$, where the inputs are binary and the addition is over real numbers. Phase detection in the probabilistic setting is fully solved by our technique. Solving the problem in the adversarial setting, however, is significantly harder, for the capacity region of the zero-error channel coding counterpart is yet unknown.

\section*{Acknowledgement}
We would like to thank Tuvi Etzion for the pointer to many relevant works and for helpful comments on an earlier version of the results. We are grateful to Ronny Roth for interesting discussions. We are also thankful to the Associate Editor and anonymous reviewers, who read the manuscript thoroughly​ and provided many helpful comments.

\bibliographystyle{IEEEtran}
\bibliography{debruijn}

\begin{IEEEbiographynophoto}{Lele Wang} 
received the B.E. degree from Tsinghua University in 2009 and the Ph.D. degree from University of California, San Diego (UCSD) in 2015, both in Electrical Engineering. She is currently a joint postdoctoral fellow at Stanford University and Tel Aviv University. Her research focus is on information theory, coding theory, and communication theory. She is a recipient of the 2013 UCSD Shannon Memorial Fellowship and the 2013--2014 Qualcomm Innovation Fellowship.
\end{IEEEbiographynophoto}

\begin{IEEEbiographynophoto}{Sihuang Hu} 
received the B.Sc. degree in 2008 from Beihang University, Beijing,
China, and the Ph.D. degree in 2014 from Zhejiang University, Hangzhou, China,
both in applied mathematics. He is currently a postdoc at Tel Aviv University.
Before that, he was a postdoc at RWTH Aachen University, Aachen, Germany,
from 2014 to 2015. 
\end{IEEEbiographynophoto}

\begin{IEEEbiographynophoto}{Ofer Shayevitz}
received the B.Sc. degree from the Technion Institute of Technology, Haifa, Israel, in 1997 and the M.Sc. and Ph.D. degrees from the Tel-Aviv University, Tel Aviv, Israel, in 2004 and 2009, respectively, all in electrical engineering. He is currently a Senior Lecturer in the Department of EE - Systems at the Tel Aviv University, and also serves as the head of the Advanced Communication Center (ACC). Before joining the department, he was a postdoctoral fellow in the Information Theory and Applications (ITA) Center at the University of California, San Diego, from 2008 to 2011, and worked as a quantitative analyst with the D. E. Shaw group in New York from 2011 to 2013. Prior to his graduate studies, he served as an engineer and team leader in the Israeli Defense Forces from 1997 to 2003, and as an algorithms engineer at CellGuide from 2003 to 2004. Dr. Shayevitz is the recipient of the ITA postdoctoral fellowship (2009 - 2011), the Adams fellowship awarded by the Israel Academy of Sciences and Humanities (2006 - 2008), the Advanced Communication Center (ACC) Feder Family award (2009), and the Weinstein prize (2006 - 2009).
\end{IEEEbiographynophoto}

\end{document}